\documentclass[twocolumn]{article}

\usepackage[utf8]{inputenc}
\usepackage[T1]{fontenc}
\usepackage{hyperref}
\usepackage{url}
\usepackage{booktabs}
\usepackage{amsfonts}
\usepackage{amsmath}
\usepackage{nicefrac}
\usepackage{microtype}
\usepackage{graphicx}
\usepackage{subcaption}
\usepackage{float}
\usepackage[super,compress,sort,nomove]{cite}  
\usepackage{geometry}
\graphicspath{ {./} {./images/} }

\usepackage[font=small,labelfont=bf]{caption}

\title{Photonic convolutional neural network with pre-trained in situ training}

\author{\large Saurabh Ranjan$^{1,*}$, Sonika Thakral$^{1}$, and Amit Sehgal$^{2}$}
\date{\vspace{-2em}} 

\begin{document}

\twocolumn[
  \begin{@twocolumnfalse}
    \maketitle
    \vspace{-1.5em}
    \begin{abstract}

Convolutional neural networks (CNNs) have transformed image processing, but the energy
consumption and inference latency of electronic based implementations remain fundamental bottlenecks.
These limitations have motivated the search for alternative hardware architectures beyond
Complementary metal-oxide-semiconductor (CMOS) chips. Optical systems can perform linear
matrix operations at the speed of light with extremely low energy dissipation, making
them attractive for CNN acceleration. However, building a fully coherent photonic CNN that
performs both linear and nonlinear operations --- and training it efficiently --- remains an
open challenge. Here we present a fully photonic convolutional neural network (PCNN) that executes image
classification in the optical domain, including convolution, max-pooling, nonlinear
activation, and fully connected layers. The network achieves $94.49\%$ accuracy on the MNIST dataset distributed across
Mach--Zehnder Interferometer (MZI) meshes, weighted Multimode Interferometer (MMI) trees, and a microring-resonator-based nonlinearity. A mathematically exact differentiable digital twin, enables backpropagation for \textit{ex situ} pre-training, reaches $97.45\%$ digital accuracy.
Trained phases are transferred one-to-one to the photonic hardware and refined via a gradient-free algorithm that
estimates the full gradient with only two forward passes. The architecture
exhibits inherent robustness to non-idealities, under
the compound effect of propagation loss, MZI insertion loss, fabrication disorder, and thermal
crosstalk. A bottom-up power analysis yields
$10.83$~W static chip consumption and $843$~ns inference latency, translating to
$220$--$330\times$ greater energy efficiency than state-of-the-art electronic GPUs for
single-image inference.

    \end{abstract}
    \vspace{2em}
  \end{@twocolumnfalse}
]

\footnotetext[1]{Shaheed Sukhdev College of Business Studies, University of Delhi, New Delhi, Delhi 110085}
\footnotetext[2]{Hansraj College, University of Delhi, New Delhi, Delhi 110007}
\footnotetext{Correspondence: saurabh.22549@sscbs.du.ac.in}

\vspace{1em}
\section{Introduction}

Convolutional neural networks (CNNs) have transformed image classification~\cite{sadeghzadeh2022}, computer vision~\cite{shen2017}, speech recognition~\cite{xiang2023}, radio signal processing~\cite{oshea}, and natural language processing~\cite{young2018}. Their computational backbone---convolution, pooling, and dense matrix products---maps almost entirely onto multiply-accumulate (MAC) operations, placing immense pressure on electronic processors~\cite{chang2018}. Graphics Processing Units (GPUs) and Tensor Processing Units (TPUs) have scaled to meet this demand, but at the cost of power budgets that now exceeds 300-700~W per chip~\cite{nvidia_h100_2022, 
jouppi2023tpuv4} and memory-bandwidth walls that limit single-inference latency.

Silicon photonics offers an alternative paradigm: coherent interference in passive waveguide meshes can execute unitary matrix--vector products at the speed of light with near-zero marginal energy per operation~\cite{shen2017, harris2018}. Shen et al.\ demonstrated a programmable nanophotonic processor implementing arbitrary $N \times N$ unitary transformations through cascaded MZI meshes~\cite{shen2017}. Subsequent work extended this to spiking networks~\cite{feldmann2019}, diffractive deep neural networks~\cite{lin2018}, photonic tensor cores~\cite{miscuglio2020, feldmann2021}, and multiply-accumulate engines~\cite{nahmias2019}. More recently, Bandyopadhyay et al.\ demonstrated single-chip photonic deep neural networks with forward-only training~\cite{bandyopadhyay2024}, and Wright et al.\ showed that deep physical networks could be trained end-to-end with backpropagation~\cite{wright2022}.

A standard CNN comprises three core operations: convolution for feature extraction, pooling for dimensionality reduction, and fully connected layers for classification~\cite{mehrabian2018, xu2019}. Among these, the pooling layer is critical for reducing spatial dimensions and model complexity~\cite{huang2023, zafar2022}. Several optical convolution implementations have been reported, including incoherent optical convolution integrated on a low-loss silicon nitride (SiN) platform~\cite{xu2019}, kernel-pruned silicon photonic designs~\cite{huang2023}, compact MMI-based units~\cite{meng2023}, AWG-based convolvers~\cite{zhang2024}, and phase-change metasurface approaches~\cite{wu2021, wei2023}. Despite these advances, most photonic neural networks still rely on frequent optical--electrical--optical (O/E/O) conversions for pooling and nonlinear activation, negating much of the latency and energy advantage of optical processing.

In this work, the PCNN realises three key advances spanning devices, architectural hardware, and algorithms:

\begin{itemize}

    \item \textbf{Depthwise--pointwise separable convolution.}
    The convolutional layer performs the convolution operation between selected input values and kernel values.
    Simultaneously reducing the spatial dimension and expanding the number of channels within the same layer
    are the challenging aspect of this operation.
    In convolutional layer~2, we therefore employ a depthwise--pointwise decomposition that uses an MZI
    binary tree followed by a CMXU mesh to reduce the spatial dimensions of the Pool~1 output and expand
    the number of channels, respectively.
    Depthwise convolution reduces the image dimension by performing the spatial convolution, while pointwise
    convolution expands the channel count from 4 to 8 for richer spatial feature representation.

    \item \textbf{WDM-based all-optical max-pooling.}
    With the expansion of channels, serial streaming of optical fields becomes a throughput bottleneck, making
    parallel streaming essential for efficient max-pooling.
    We therefore implement a wavelength-division multiplexing architecture in which multiple feature channels
    share a single physical delay network through an Arrayed Waveguide Grating (AWG), achieving max-pooling
    without any electronic conversion~\cite{amiri2025}.
    This reduces the pooling hardware footprint from $\mathcal{O}(N)$ to $\mathcal{O}(1)$ in the dominant
    delay-line area, meaning that our WDM-based pooling scales at constant hardware cost for $N \geq 8$,
    where $N$ is the number of channels.

    \item \textbf{Hybrid \textit{ex situ/ in situ} training.}
    Training an optical neural network efficiently is one of the hardest problems in photonic computing.
    The critical bottleneck lies in the forward inference, which requires repeated evaluation of the entire
    training set to estimate gradients, and in backpropagation, which propagates gradient values backwards
    to adjust weights and minimise the loss.
    To address this, a differentiable digital twin pre-trains the model using backpropagation, after which the trained values are transferred directly to the photonic
    hardware, then hardware non-idealities are introduced to simulate realistic on-chip
    conditions.
    \textit{In situ} fine-tuning via Simultaneous Perturbation Stochastic Approximation
    (SPSA)~\cite{spall1992} then compensates for the non-idealities using only two forward
    passes per gradient estimate.
    Running for 500 steps, this fully optical \textit{in situ} fine-tuning raises the model's test accuracy
    to $94.49\%$ on the MNIST test set, establishing this hybrid training method among the powerful
    and accurate training approaches for photonic CNN architectures, presented till date.

\end{itemize}

This PCNN system, comprising 1,796 individually tunable weight parameters, is one of the largest
photonic CNN chips to date in terms of the number of on-chip weights.
To the best of our knowledge, this is the first demonstration of a photonic CNN that combines all
subsystems---convolution, pooling, fully connected layers, and nonlinear activation---into a single
photonic integrated circuit (PIC), performing all convolutional, pooling, linear, and nonlinear
computations entirely in the optical domain, achieving $94.49\%$ final accuracy on MNIST, and
implementable on an $18 \times 18\ \text{mm}^2$ chip.

\begin{figure*}[t]
  \centering
  \includegraphics[width=0.95\textwidth]{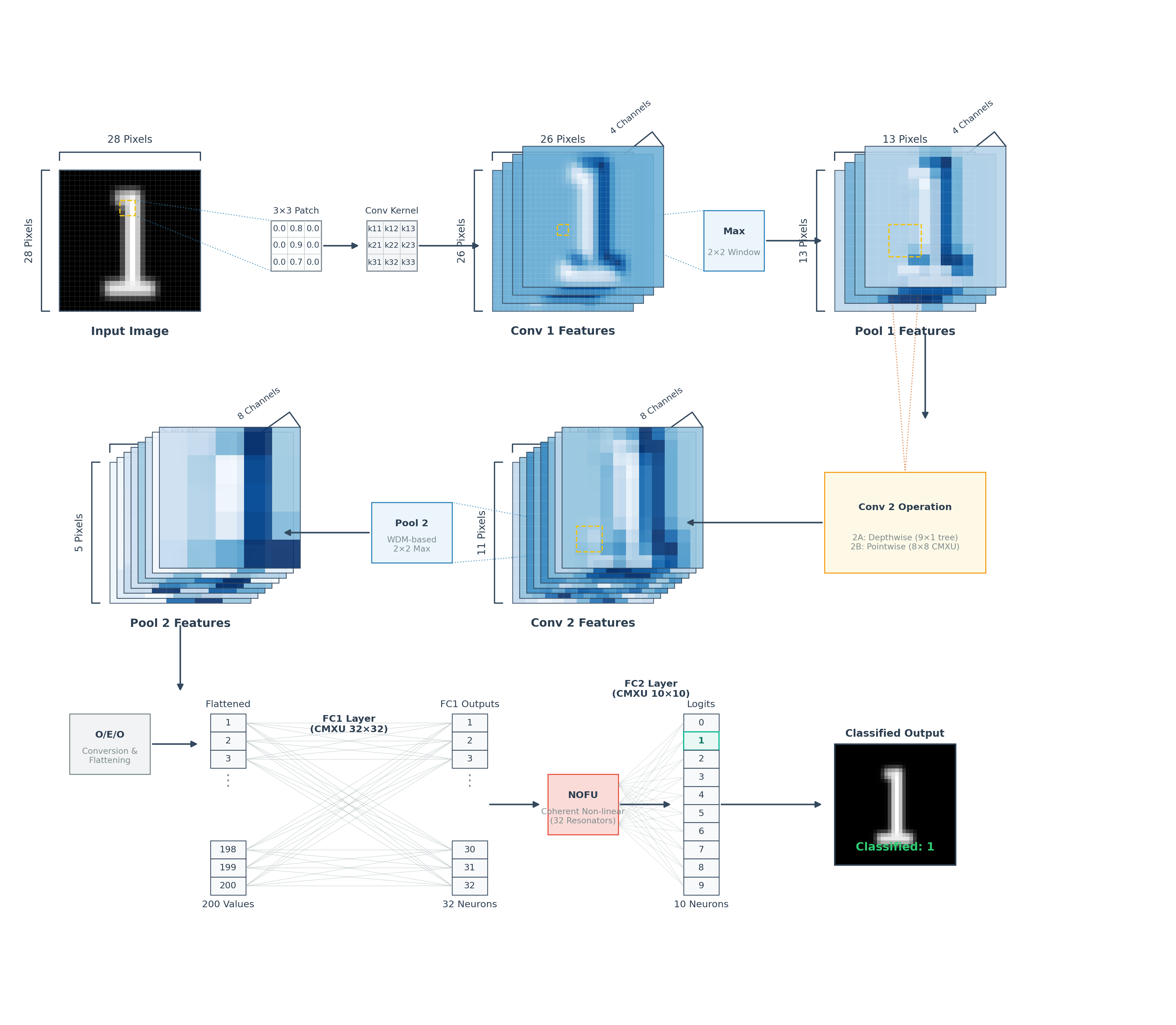}
  \caption{\textbf{Block diagram of the PCNN and its operation in MNIST data classification.} A $28 \times 28$ input image of digit ``1'' is amplitude-modulated at 1~GHz and propagated through the photonic pipeline. Conv1 extracts 4 feature channels ($26 \times 26$) via MZI mesh processing; Pool1 reduces spatial dimensions to $13 \times 13$ using all-optical MMI+GST comparators; Conv2 deepens to 8 channels ($11 \times 11$) via depthwise separable MZI trees (2A) and pointwise CMXUs (2B); Pool2 applies WDM-multiplexed max-pooling ($5 \times 5$); FC1 + NOFU produce 32 nonlinearly activated features via microring resonators; FC2 maps to 10 output probabilities, correctly predicting digit 1 with high confidence.}
  \label{fig:dataflow}
  \vspace{0.5em}\hrule
\end{figure*}

\section{Architecture overview}

The PCNN maps a standard CNN topology onto silicon photonic hardware, replacing transistor-based arithmetic with optical interference and passive delay networks. Figure~\ref{fig:dataflow} summarises the end-to-end dataflow. The architecture comprises seven photonic stages processing a $28 \times 28$ greyscale handwritten digits image into a 10-class probability vector. It consists of convolutional layer, pooling layer, fully connected layer and nonlinearity layer.


\subsection{Image encoding}

The authors utilised a serial Mach–Zehnder modulator (MZM)~\cite{paula2023} which converts the input image into an amplitude-modulated optical stream. Each MNIST image is a $28 \times 28$ greyscale image, giving 784 pixels with raw integer intensities between $[0, 255]$. Before modulation, these values are normalised to the unit
interval: $I_{ij} = p_{ij} / 255$, where $p_{ij} \in \{0, \ldots, 255\}$
is the raw pixel value at row $i$, column $j$, converting $I_{ij}$ between $[0, 1]$. 

The MZM encodes each normalised pixel as an amplitude-modulated optical
field consists of both amplitude and phase. A continuous-wave (CW) laser feeds one input arm of the MZI-based modulator and the other arm will get equal and opposite voltage.  
By applying a drive voltage proportional to $I_{ij}$ to the
thermo-optic phase shifter inside the MZM, the interference condition at
the output coupler is set so that the transmitted field amplitude scales
as $\sqrt{I_{ij}}$. The output optical field for pixel $(i,j)$ is therefore
\begin{equation}
  E_{ij} = \sqrt{I_{ij}}\,e^{j\phi_0},
  \label{eq:encoding}
\end{equation}
where $\phi_0$ is the fixed carrier phase set by the laser. The optical
intensity at the modulator output is $|E_{ij}|^2 = I_{ij}$, which recovers
the original normalised pixel value at any photodetector downstream. 

Crucially, the field $E_{ij}$ is a complex quantity: it carries both an
amplitude ($\sqrt{I_{ij}}$) and a phase ($\phi_0$). This is precisely what
the Clements MZI mesh downstream operates on---complex optical amplitudes,
and it enables the mesh to realise a
unitary transformation.

The 784 pixels are serialised in raster-scan order (row-major) and fed
into the MZM at a 1~GHz pixel clock, producing 784 consecutive optical
pulses, each of duration 1~ns, on a single waveguide. This serial stream
is then routed into the tapped delay line that feeds into Conv1.

\subsection{Convolutional layer~1}

Conv1 extracts $3\times3$ spatial features from a $28\times28$ greyscale image using a streaming architecture. A $1\times9$ MMI splitter fans the serial pixel stream into 9 parallel waveguides, each routed through a $\text{Si}_3\text{N}_4$ spiral delay line. The delay assigned to each path is chosen so that the earlier-arriving pixels are held longer, ensuring all 9 pixels of a $3\times3$ patch exit their respective delay lines simultaneously.

The first $3\times3$ patch spans the raster positions $(0,0)$ through $(2,2)$: pixel $E_{00}$ arrives at clock cycle $t_0$ and is delayed by 58~ns; $E_{01}$ arrives at $t_1$, delayed 57~ns; $E_{02}$ at $t_2$, delayed 56~ns; $E_{10}$ at $t_{28}$, delayed 30~ns; $E_{11}$ at $t_{29}$, delayed 29~ns; $E_{12}$ at $t_{30}$, delayed 28~ns; $E_{20}$ at $t_{56}$, delayed 2~ns; $E_{21}$ at $t_{57}$, delayed 1~ns; and $E_{22}$ at $t_{58}$, with zero delay, since it is the last pixel to arrive. All nine fields are therefore released simultaneously at $t = 58$~ns, forming a synchronised patch. This patch is zero-padded to a 10-element complex column vector:
\begin{multline}
  \mathbf{v} = \bigl[E_{00},\, E_{01},\, E_{02},\, E_{10},\, E_{11},\, E_{12},\, E_{20},\, E_{21},\, E_{22},\, 0\bigr]^\top \\
  \in \mathbb{C}^{10}.
  \label{eq:patch_vec}
\end{multline}

With a $28\times28$ input and stride 1, this process yields $P_1 = 26\times26 = 676$ overlapping patches in total.

\begin{figure*}[t]
  \centering
  \includegraphics[width=0.98\textwidth]{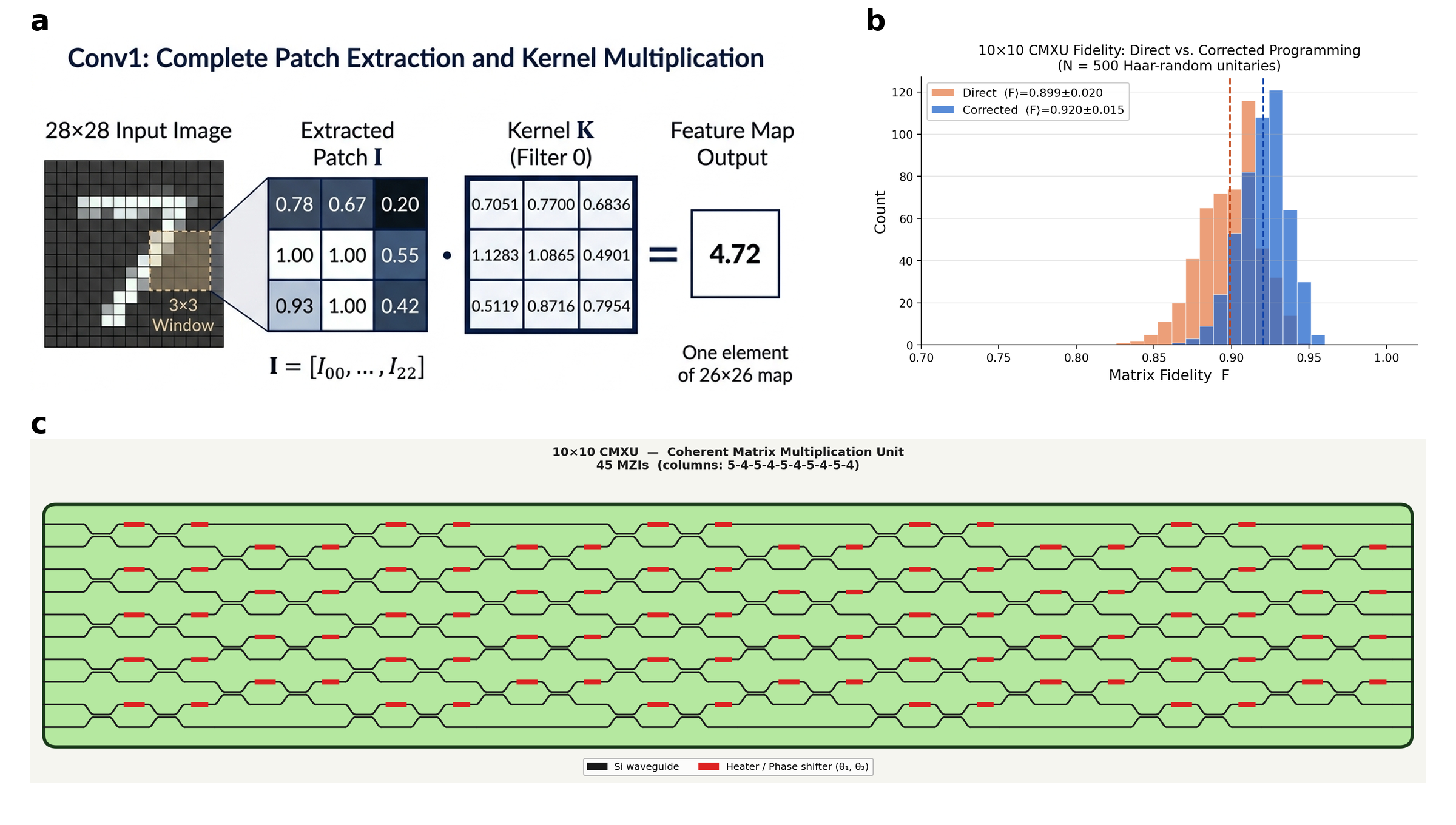}
  \caption{\textbf{Conv1 patch extraction flow, matrix fidelity, and physical layout.}
    (a)~Conv1 patch extraction and kernel multiplication flow: a $3\times3$ sliding window traverses the $28\times28$ input image in raster order at 1~GHz, yielding 676 overlapping patches. Each patch is multiplied element-wise with Kernel~0 to produce Feature Map 0.
    (b)~$10\times10$ CMXU matrix fidelity distribution over 500 Haar-random unitary matrices under direct programming (orange, $\langle F\rangle = 0.899$) and SPSA-corrected programming (blue, $\langle F\rangle = 0.920$). Both distributions assume $\sigma_\theta = 0.04$~rad phase noise and nearest-neighbour crosstalk $\xi = 0.05$.
    (c)~Layout schematic of the $10\times10$ Coherent Matrix Multiplication Unit (CMXU), consisting of 45 MZIs configured in a symmetric Clements mesh topology split across 10 layers.}
  \label{fig:conv1_combined}
\end{figure*}

The complete patch extraction and vector--matrix multiplication process are illustrated in Fig.~\ref{fig:conv1_combined}(a). Each $3\times3$ patch is flattened into a row vector $\mathbf{I} = [I_{00}, I_{01}, I_{02}, I_{10}, I_{11}, I_{12}, I_{20}, I_{21}, I_{22}]$ and multiplied element-wise with the kernel $\mathbf{K}$, whose trained values for Filter~0 are
\begin{equation}
  \mathbf{K}^{(0)} =
  \begin{bmatrix}
    0.7051 & 0.7700 & 0.6836 \\
    1.1283 & 1.0865 & 0.4901 \\
    0.5119 & 0.8716 & 0.7954
  \end{bmatrix},
  \label{eq:kernel}
\end{equation}
yielding the feature-map activation $A = \mathbf{I} \cdot \mathbf{K} = \sum_{m,n} I_{mn} K_{mn}$, one scalar per patch position.

In our PCNN, the convolutional kernel is not stored as an explicit $3\times3$ matrix of scalar weights, as it would be in a standard digital CNN. Instead, the weights are an emergent property of the optical interference inside the Clements MZI mesh. The effective $3\times3$ spatial filter for Filter~0 corresponds to the first 9 elements of the first row of the learned $10\times10$ unitary matrix $U$. The phase angles $\theta_1$ and $\theta_2$ of each MZI are trained directly, and the $3\times3$ kernel weights arise implicitly from the resulting $U$.

Each patch vector $\mathbf{v}_{r,c}$ propagates through a $10\times10$ Coherent Matrix Multiplication Unit (CMXU)\cite{bandyopadhyay2024}---a rectangular Clements mesh~\cite{clements2016} of 45 MZIs implementing an arbitrary $10\times10$ unitary transformation (Fig.~\ref{fig:conv1_combined}(c)). The mesh has 100 tunable phase parameters, including 10 external output phases. Since the network is trained by optimising the phase angles, it learns to concentrate discriminative features into the first 4 output ports; the remaining 6 ports provide over-parameterisation that gives the optimiser additional degrees of freedom to shape the feature projection. The outputs of these 4 ports are accumulated over all 676 time steps, yielding four $26\times26$ feature maps. Each MZI consists of two directional couplers (DCs), an internal phase shifter $\theta_1$ between the two couplers, and an external phase shifter $\theta_2$ on the output arm.

Because $\det(U) = 1$, optical power is conserved (except for MZI insertion loss). Training with unitary layers avoids the vanishing-gradient problem, improving optimisation stability~\cite{jing2017}.

To validate the $10\times10$ CMXU, we programmed 500 Haar-random unitary targets into the simulated device and evaluated the normalized trace fidelity $F = |\mathrm{Tr}[U_{\mathrm{target}}^\dagger U_{\mathrm{sim}}]| / 10$ as the accuracy metric. Two scenarios were evaluated: (i)~\emph{direct programming}, in which the Clements decomposition provides nominal phase angles subsequently perturbed by Gaussian calibration noise and thermal crosstalk, yielding $\langle F\rangle = 0.899 \pm 0.020$; and (ii)~\emph{corrected programming}, in which gradient-free SPSA phase correction is applied, achieving $\langle F\rangle = 0.920 \pm 0.015$ (Fig.~\ref{fig:conv1_combined}(b)). The residual gap from the ideal fidelity ($F = 1.0$) reflects the known limitation that gradient-free optimisers converge slowly in high-dimensional landscapes and would benefit from extended fine-tuning.

\subsection{Pooling layer~1}

Max-pooling is a non-linear down-sampling operation that reduces the spatial dimensions of a feature map by selecting the maximum value within each local window, thereby retaining the most salient activations while discarding redundant spatial information. Implementing this operation in the optical domain is a key challenge for a fully photonic CNN, since it requires a nonlinear comparison mechanism without recourse to electronic detection. Pool1 in our PCNN performs $2\times2$ max-pooling on each of the 4 output channels from Conv1, reducing spatial resolution from $26\times26$ to $13\times13$ entirely within the optical domain.

Because pixels arrive sequentially in raster-scan order at a 1~GHz clock rate, the four pixels belonging to any $2\times2$ spatial window enter the pooling layer at different times. A $1\times4$ MMI splitter first fans out the incoming optical field into four parallel waveguides, each routed through a silicon nitride ($\text{Si}_3\text{N}_4$) spiral delay line. For a feature map of width $W=26$, the delays required to bring all four pixels of a $2\times2$ window into simultaneous temporal alignment are $\Delta t = [27,\,26,\,1,\,0]$~ns, so that the top-left pixel $(0,0)$---which arrived earliest---is held in the longest delay line until the bottom-right pixel $(1,1)$ arrives at the zero-delay tap. All four aligned fields are then passed to the wavelength conversion and comparison stages.

The design of pooling unit is taken directly from
Amiri et al.~\cite{amiri2025}, who first demonstrated this concept on all-optical CNN. 
To enable all-optical comparison, the aligned fields are duplicated into a data path and a control path. The data path retains the original carrier wavelength $\lambda_0 = 1553.6$~nm. The control path is fed into a four-wave mixing (FWM) wavelength converter---a highly nonlinear silicon waveguide pumped by an auxiliary continuous-wave (CW) laser---which up-converts the control copies to $\lambda_1 = 1568.7$~nm while preserving relative optical amplitudes. The two wavelengths are thus spectrally separated, permitting independent downstream processing without crosstalk.

\begin{figure*}[t]
  \centering
  \includegraphics[width=0.98\textwidth]{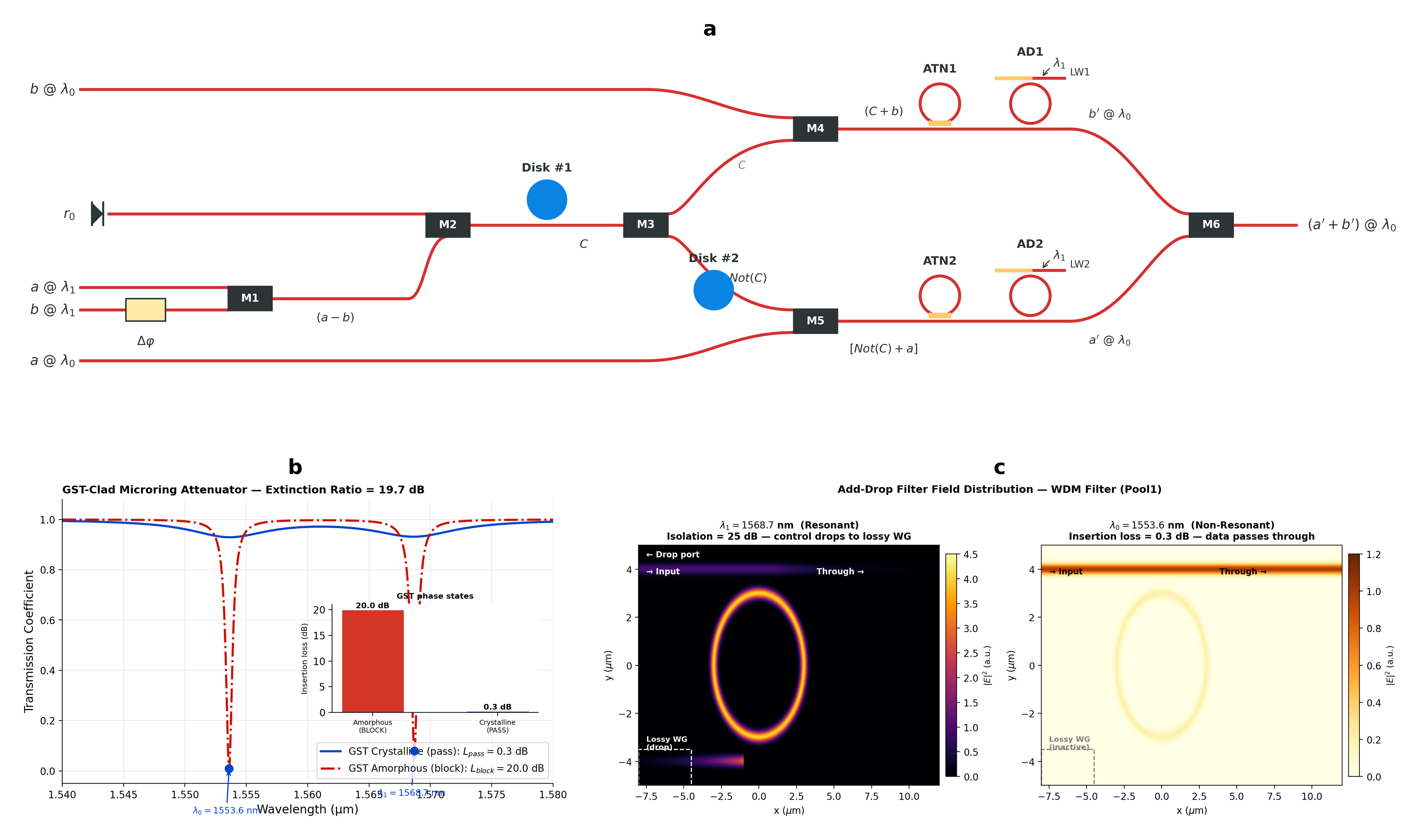}
  \caption{\textbf{All-optical Pool1 max-pooling hardware.}
    (a)~Schematic of the Optical Max Unit (OMU): two data signals $a$ and $b$ (at $\lambda_0$) are compared in the power domain. Their control copies at $\lambda_1$ are subtracted via M1 and the difference is injected into a bistable Graphene-on-Silicon microring (Disk~\#1) biased by a reference signal $r_0$. The comparator output $C$ is inverted by an all-optical NOT gate (Disk~\#2) to yield $\overline{C}$; these complementary signals gate GST-clad attenuators ATN1 and ATN2, selecting the larger field, while add-drop filters AD1 and AD2 remove the $\lambda_1$ control tone to deliver the winner exclusively at $\lambda_0$.
    (b)~Lorentzian through-port transmission of the GST-clad microring attenuator for the amorphous (blocking, $L_\text{block}=20.0$~dB) and crystalline (passing, $L_\text{pass}=0.3$~dB) phase states, yielding a $19.7$~dB extinction ratio.
    (c)~Simulated optical intensity $|E|^2$ in the WDM add-drop microring filter: at $\lambda_1=1568.7$~nm (left) the field couples into the ring and exits the drop port with $25$~dB isolation; at $\lambda_0=1553.6$~nm (right) the bus waveguide is off-resonance and the field passes with only $0.3$~dB loss.}
  \label{fig:pool1_combined}
\end{figure*}

The actual max selection is performed by a binary tree of three Optical Max Units (OMUs)~\cite{amiri2025}: Stage~1 evaluates $\text{OMU}_1 = \max(p_0, p_1)$ and $\text{OMU}_2 = \max(p_2, p_3)$ in parallel, and Stage~2 evaluates $\text{OMU}_3 = \max(\text{OMU}_1, \text{OMU}_2)$ to yield the pool output. Within each OMU, the two control copies $a$ and $b$ at $\lambda_1$ are subtracted via a $2\times2$ MMI coupler and the resulting difference field is injected into a Graphene-on-Silicon bistable microring resonator (Disk~\#1) biased by a reference continuous-wave signal $r_0$ at $\lambda_1$ whose intensity $|r_0|^2 = P_0$ sets the switching threshold. If $P_a > P_b$ the injected power exceeds $P_0$ and the disk switches to its low-transmission state, asserting $C = \text{LOW}$; if $P_a < P_b$ the disk remains in its high-transmission bistable state, asserting $C = \text{HIGH}$. A secondary graphene microdisk (Disk~\#2) acting as an all-optical NOT gate inverts $C$ to produce the complementary signal $\overline{C}$, ensuring that exactly one of the two data-path signals is passed and the other is blocked (Fig.~\ref{fig:pool1_combined}(a)).

The gate signals $C$ and $\overline{C}$ at $\lambda_1$ drive a pair of $\text{Ge}_2\text{Sb}_2\text{Te}_5$ (GST) phase-change microring attenuators~\cite{amiri2025,feldmann2019,wu2021}. When a GST ring receives a HIGH control signal it transitions to its crystalline phase, leaving the bus waveguide nearly unperturbed ($L_\text{pass} \approx 0.3$~dB, Fig.~\ref{fig:pool1_combined}(b)); a LOW control signal keeps the ring in its amorphous phase, strongly coupling light into the lossy resonator and blocking the data signal with an extinction ratio of $\sim$20~dB ($L_\text{block} \approx 20.0$~dB). Attenuator~ATN1 (carrying data $b$ at $\lambda_0$) is gated by $C$, while Attenuator~ATN2 (carrying data $a$ at $\lambda_0$) is gated by $\overline{C}$. Consequently, when $P_a > P_b$, ATN2 passes $a$ and ATN1 blocks $b$; when $P_a < P_b$, ATN1 passes $b$ and ATN2 blocks $a$. The two attenuator outputs are recombined in a final $2\times2$ MMI coupler and the waveguide then contains the winner field at $\lambda_0$ together with residual control power at $\lambda_1$.

To isolate the pooled output, the combined waveguide is routed through a WDM add-drop microring filter resonant at $\lambda_1$, which evanescently couples the residual control signal into a lossy drop port with $\sim$25~dB isolation, while leaving $\lambda_0$ off-resonance with only 0.3~dB insertion loss (Fig.~\ref{fig:pool1_combined}(c)). The final pool output thus emerges as a clean, single-wavelength coherent field at $\lambda_0$, ready to be streamed into the second convolutional layer.

\subsection{Convolutional layer~2}

\begin{figure*}[t]
  \centering
  \includegraphics[width=0.95\textwidth]{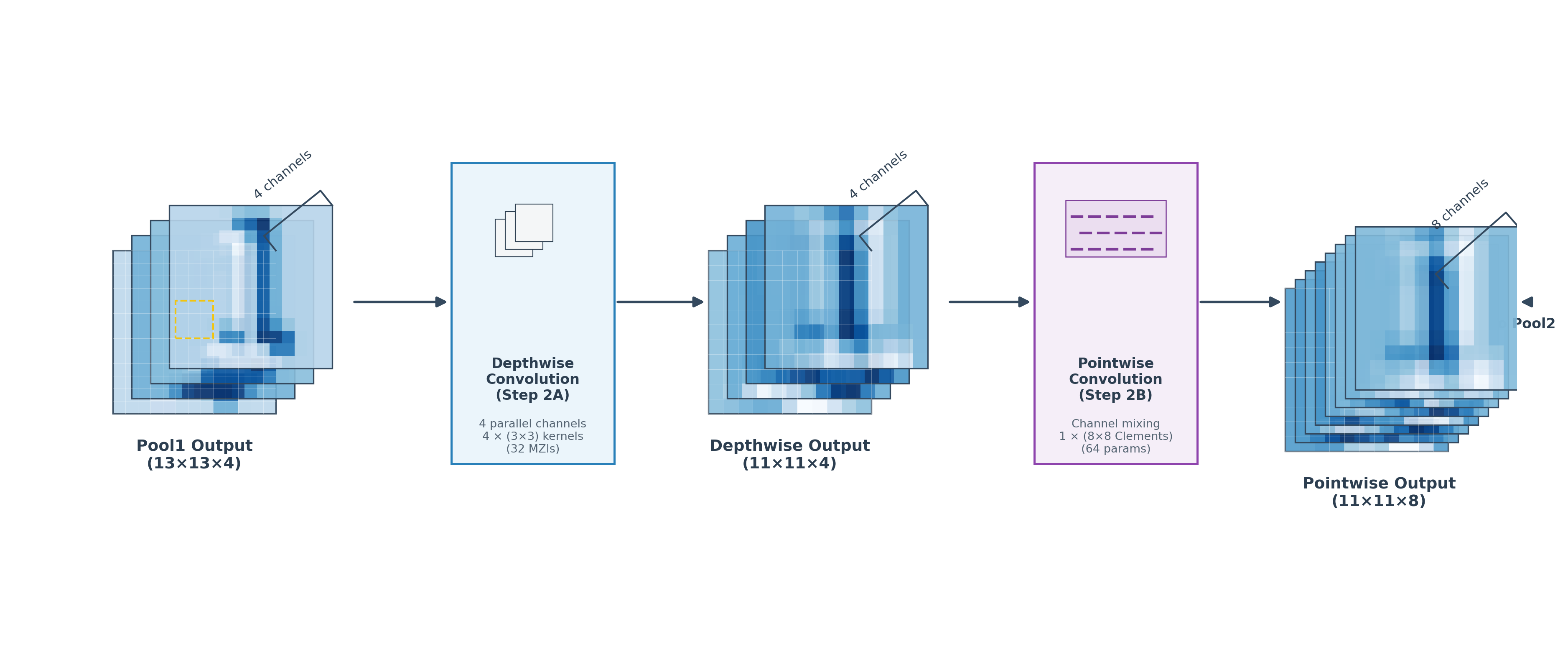}
  \caption{\textbf{Architectural dataflow and pipeline of Convolutional Layer~2.}
  The $13\times13\times4$ feature maps from Pool1 are processed via depthwise-separable convolution. 
  First, in Step~2A (Depthwise Convolution), four parallel $3\times3$ kernels (each implemented with a 9-tap delay line and a $9\rightarrow1$ MZI combiner tree, for a total of 32 MZIs) perform spatial filtering on each channel independently, yielding a $11\times11\times4$ intermediate volume. 
  Second, in Step~2B (Pointwise Convolution), a single $8\times8$ Clements mesh (64 parameters) mixes the 4-channel intermediate fields across channels, yielding the final $11\times11\times8$ feature maps for input to Pool2.}
  \label{fig:conv2_working}
\end{figure*}

The second convolutional layer receives the $13\times13\times4$ optical feature map from Pool1 and produces 8 output channels of $11\times11$ spatial resolution each, using a depth-point separable factorisation. This decomposition splits a conventional $3\times3$ convolution over 4 input channels into two sequential steps: Step~2A performs per-channel spatial filtering (depthwise), and Step~2B performs cross-channel mixing (pointwise). This factorisation reduces the Conv2 hardware to a total of 128 trainable parameters. The overall architectural dataflow of Convolutional Layer~2 is illustrated in Fig.~\ref{fig:conv2_working}.

\subsubsection{Step~2A: Depthwise Convolution}

Each of the 4 input channels from Pool1 arrives as a serial stream of 169 optical field samples $E_{ij}$ at 1~GHz, representing the $13\times13$ feature map in raster-scan order. All four channels are processed identically and in parallel, each through an independent hardware pipeline consisting of a $1\times9$ MMI splitter, a 9-tap delay network, and a $9\rightarrow1$ MZI binary combiner tree.

To form a $3\times3$ convolution patch, the 9 pixels of the patch must be presented simultaneously to the combiner tree despite arriving at 9 different time steps. A $1\times9$ MMI splitter fans out the serial optical stream into 9 parallel waveguides, each routed through a spiral delay line of a different length. The delay assigned to each path is chosen so that pixels arriving earlier in the raster stream undergo a longer delay, and those arriving later undergo a shorter delay, ensuring all 9 are released simultaneously. For the $13\times13$ input (width $W = 13$), the 9 delay values in nanoseconds are $\Delta t = [28, 27, 26, 15, 14, 13, 2, 1, 0]$~ns, assigned to paths 0 through 8 respectively (detailed timing derivation in Supplementary Section~S2). At $t = 28$~ns, all 9 pixels $\{I_{00}, I_{01}, \ldots, I_{22}\}$ exit their respective delay lines simultaneously, forming the first patch centred at $[1,1]$. The sliding window advances by one pixel at each subsequent clock cycle, and with stride~1 on a $13\times13$ input, the depthwise stage produces $P_2 = 11\times11 = 121$ output patches per channel( $P_1 = 676$ patches in Conv1).

We employ a binary MZI combiner tree that coherently combines the 9 simultaneously presented patch samples into a single output field. As Miller demonstrates, the binary MZI tree is the shortest-path structure for coherently combining an arbitrary spatial vector field into a single waveguide~\cite{Miller2013, Miller2020}. In this combiner tree, each MZI has two inputs and two outputs; we always select the \emph{top} output port and discard the bottom port. The 9 patch inputs $x_0$ through $x_8$ are combined across 4 tree levels using 8 MZIs (16 tunable phase parameters):

\begin{align}
  \text{Level~1 (4 MZIs):} &\quad y_k = U_k(x_{2k},\, x_{2k+1}), \quad k = 0,1,2,3 \\[4pt]
  \text{Level~2 (2 MZIs):} &\quad z_0 = U_4(y_0, y_1), \\
                            &\quad z_1 = U_5(y_2, y_3) \\[4pt]
  \text{Level~3 (1 MZI):}  &\quad w_0 = U_6(z_0, z_1) \\[4pt]
  \text{Level~4 (1 MZI):}  &\quad \text{Output} = U_7(w_0, x_8)
\end{align}

where $x_8$ bypasses the upper tree via a direct delay path, and each $U_k$ is the $2\times2$ MZI unitary parameterised by its internal phase $\theta_{1,k}$ (setting the split ratio) and external phase $\theta_{2,k}$ (controlling the output relative phase):

\begin{equation}
  U_k(\theta_{1,k}, \theta_{2,k}) = i\,e^{i\theta_{1,k}/2}
  \begin{pmatrix}
    e^{i\theta_{2,k}}\sin\!\tfrac{\theta_{1,k}}{2} & e^{i\theta_{2,k}}\cos\!\tfrac{\theta_{1,k}}{2} \\
    \cos\tfrac{\theta_{1,k}}{2} & -\sin\tfrac{\theta_{1,k}}{2}
  \end{pmatrix}.
  \label{eq:mzi_tree_unit}
\end{equation}

The combined output can be written as a weighted inner product of the input vector:
\begin{equation}
  y_{\text{out}} = \mathbf{w}^\dagger \mathbf{x}, \qquad \mathbf{x} = [x_0, \ldots, x_8]^\top \in \mathbb{C}^9,
\end{equation}
where the effective weight vector $\mathbf{w} \in \mathbb{C}^9$ is determined by the 16 phase parameters $\{\theta_{1,k}, \theta_{2,k}\}_{k=0}^{7}$. By tuning these parameters, the tree can represent any complex unit-norm weight vector subject only to physical power conservation ($\|\mathbf{w}\|_2 \leq 1$), providing full expressiveness for a single spatial filter. The weight vector $\mathbf{w}$ is therefore exactly equivalent to a $3\times3$ convolutional kernel, and the parameters $\{\theta_{1,k}, \theta_{2,k}\}$ are trained offline via \textit{ex situ} backpropagation and \textit{in situ} fine-tuning (described in Section~\ref{sec:training}). Across 4 parallel depthwise channels, the depthwise stage totals 32 MZIs and 64 trainable parameters.

\subsubsection{Step~2B: Pointwise Convolution}

At each spatial position $(i,j)$ of the $11\times11$ feature map, the four 
depthwise outputs from Step~2A form a 4-element complex vector 
$\mathbf{v}_{ij} = [y^{(0)}_{ij},\, y^{(1)}_{ij},\, y^{(2)}_{ij},\, 
y^{(3)}_{ij}]^\top \in \mathbb{C}^4$, encoding the per-channel spatial 
response at that position. The pointwise step mixes these four channels to 
produce 8 output channels, realising a $1\times1$ convolution that 
recombines cross-channel information without any additional spatial filtering. 
The vector $\mathbf{v}_{ij}$ is zero-padded to 8 elements and fed directly 
into a single $8\times8$ Clements mesh (28~MZIs, 64 trainable phase 
parameters), implementing an arbitrary $8\times8$ unitary transformation 
$U^{(8)}$:
\begin{equation}
  \mathbf{u}_{ij} = U^{(8)} \begin{pmatrix} \mathbf{v}_{ij} \\ \mathbf{0}_4 
  \end{pmatrix}, \qquad \mathbf{u}_{ij} \in \mathbb{C}^8.
  \label{eq:pointwise}
\end{equation}
The 8 output channels correspond to the 8 rows of $U^{(8)}$, each 
representing a learned linear combination of the 4 depthwise input channels. 
Since all four depthwise channels operate in parallel with identical timing, 
$\mathbf{v}_{ij}$ is assembled simultaneously the moment position $(i,j)$ 
exits each MZI tree, and immediately forwarded to the $8\times8$ CMXU without 
any intermediate buffering.

The two steps are fully pipelined: Step~2B does not wait for Step~2A to 
complete all 121 patches. At $t = 28$~ns (local to Conv2), the first patch 
exits the MZI trees and enters the $8\times8$ CMXU; at $t = 29$~ns the CMXU 
produces its first 8-channel output while Step~2A concurrently delivers the 
next patch. The final $11\times11\times8$ output feature volume completes at 
$t = 149$~ns, giving a total Conv2 latency of 149~ns 
(see Supplementary Information SI). Across both steps, Conv2 comprises 128 trainable parameters: 
64 from the four depthwise MZI trees and 64 from the pointwise CMXU.

\subsection{Pooling layer~2}
Pool~2 reduces the $11\times11\times8$ feature volume from Conv2 to $5\times5\times8$ via $2\times2$ max-pooling applied independently across all 8 channels. Rather than replicating the full delay-network and wavelength-conversion hardware for each channel separately (as in Pool~1), Pool~2 exploits Wavelength Division Multiplexing (WDM)~\cite{zou2022} to share a single physical delay network across all 8 channels simultaneously, yielding a substantial reduction in silicon area.

\begin{figure*}[t]
  \centering
  \includegraphics[width=0.98\textwidth]{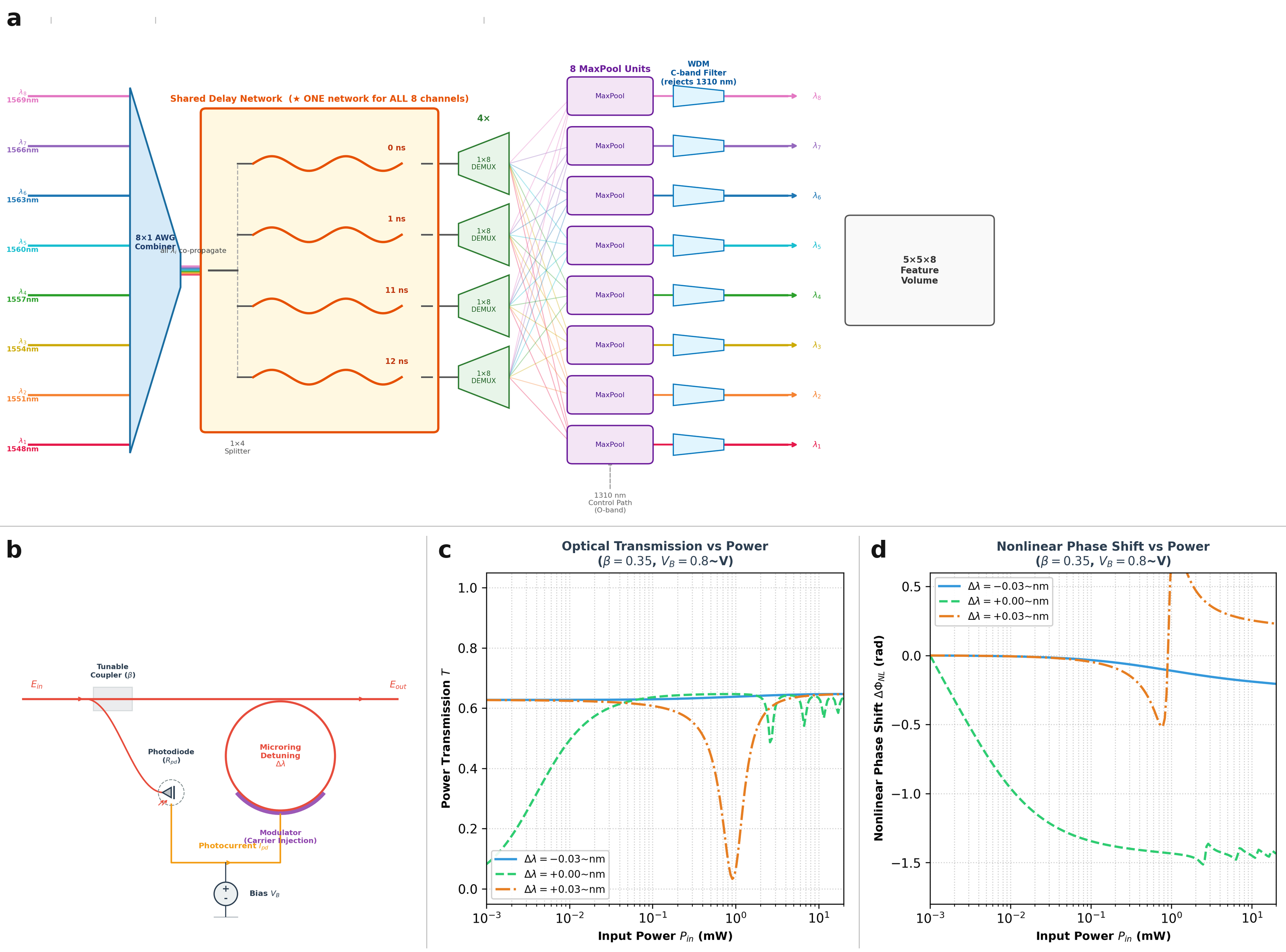}
  \caption{\textbf{WDM Pooling Layer~2 and Nonlinear Optical Function Unit (NOFU).}
    (a)~WDM Pool~2 architecture: eight C-band wavelengths $\lambda_1$--$\lambda_8$
    are combined via an $8\times1$ AWG multiplexer into a single shared delay
    network (4 spiral delay lines). A $1\times8$ AWG demultiplexer bank separates
    the aligned outputs to individual lanes; eight OMU-based max-pool units compare
    values using 1310~nm O-band control signals (crosstalk-free), and WDM bandpass
    filters remove the control path, delivering the $5\times5\times8$ feature volume.
    (b)~Optoelectronic feedback schematic of one NOFU unit: a tunable directional
    coupler (tap fraction $\beta$) routes a fraction of the input field to a
    photodiode ($R_\text{pd}$), whose photocurrent is injected under bias $V_B$
    into the carrier-injection section of a high-Q microring (detuning
    $\Delta\lambda$), nonlinearly modifying the through-port output $E_\text{out}$.
    (c)~Simulated power transmission $T$ vs.\ input power $P_\text{in}$
    ($\beta=0.35$, $V_B=0.8$~V, three detunings).
    All curves begin near $T\approx0.63$; the on-resonance and $+0.03$~nm curves
    sweep through a deep minimum as carrier injection shifts the resonance through
    the operating wavelength.
    (d)~Corresponding nonlinear phase shift $\Delta\Phi_\text{NL}$, accumulating
    a negative shift reaching ${\sim}{-1.5}$~rad at the resonance crossing.}
  \label{fig:pool2_nofu_combined}
\end{figure*}

Each of the 8 output channels is assigned a dedicated C-band wavelength at 3~nm channel spacing, $\lambda_i = 1548 + 3(i-1)$~nm for $i = 1,\ldots,8$, spanning 1548--1569~nm within the ITU C-band. An $8\times1$ AWG multiplexer combines all 8 channel signals into a single shared waveguide, which then feeds a $1\times4$ splitter and four spiral delay lines generating delays of $\{0, 1, 11, 12\}$~ns. Every wavelength traverses the same spiral network simultaneously, so all 8 channels accumulate the four temporally aligned values at no additional hardware cost. A $1\times8$ AWG demultiplexers then separates the four aligned multi-wavelength outputs back into individual $\lambda_i$ lanes, yielding $8\times4$ aligned optical signals ready for max-pooling.

Eight dedicated max-pool units — one per channel — each receives four aligned signals at $\lambda_i$ (data path at C-band) together with control path at 1310~nm (O-band). Using the O-band for control signals eliminates any crosstalk risk with the C-band data path. Each unit performs $2\times2$ max-pooling in the same binary-tree order as Pool~1~\cite{amiri2025}. A final WDM bandpass filter at each unit output transmits only the C-band data wavelengths $\lambda_1$--$\lambda_8$ and rejects the 1310~nm control signals, producing the $5\times5\times8$ feature maps forwarded to the fully connected stage.

This WDM based approach helps in reducing the delay-network cost across all 8 channels. It replaces the $\mathcal{O}(N)$ area scaling of per-channel architecture with near-constant growth. At $N = 8$, this design delivers a 76\% reduction in silicon area compared to Pool 1 style replication, which decreases the area from $\sim$81~mm$^2$ to $\sim$19~mm$^2$ (Supplementary Section III).

\subsection{Fully Connected layer 1 (FC1)}

Pool~2 outputs $5\times5\times8 = 200$ optical values arriving as 25 serial
samples per nanosecond across 8 wavelengths.  The FC~1 stage requires
all 200~values simultaneously — so we implement a opto-electronic-optical (O/E/O) bridge that converts the serial temporal
stream into 200 optical signals in three steps.

First, eight Ge-on-Si photodetectors (one per wavelength channel) detect the
incoming optical fields and convert them to photocurrent, which a
transimpedance amplifier (TIA, 1~k$\Omega$ gain) turns into a 0--1~V analog
voltage.  A 10-bit ADC (1~GS/s) samples each channel once per nanosecond,
delivering one digital value per channel per clock cycle. Second, the 25 digital
values from each of the 8~channels are shifted, one per cycle, into a bank
of 8 shift-register arrays (25 stages, 10-bit each), so by $t = 354$~ns all
$200\times10 = 2000$ flip-flops are filled and the HOLD signal fires.  Third, all 200~DACs convert their stored values to 0--1~V drive
voltages simultaneously, and 200~Mach--Zehnder modulators (MZMs, arranged
$20\times10$, fed by a shared 50~mW CW laser at $\lambda_0 = 1550$~nm)
produce 200~parallel optical signals in a single clock cycle.  The entire
serial-to-parallel conversion takes 26~ns (Supplementary Section~I).

The 200 parallel optical signals enter a three-level binary MMI tree that
reduces them to 32.  Each node is a trainable $2\times1$ MMI with
thermo-optic phase shifters on both input arms.  The relative phase
$\Delta\phi = \phi_1 - \phi_2$ controls the effective mixing weight between
the two inputs:
\begin{equation}
  |E_\text{out}|^2 = \frac{|E_1|^2 + |E_2|^2 + 2|E_1||E_2|\cos(\Delta\phi)}{2}.
\end{equation}
Level~1 reduces 200$\rightarrow$100 (100 MMIs, 200 phase parameters), Level~2 reduces
100$\rightarrow$50 (50 MMIs, 100 parameters), and Level~3 reduces 50$\rightarrow$32 using 18~MMIs
on 36~of the 50 signals while the remaining 14 pass through unchanged
(36 parameters).

The 32 optical outputs of the MMI tree are injected into a $32\times32$
rectangular Clements mesh — a fully-connected photonic linear layer.  The
mesh contains $(32\times31)/2 = 496$ MZIs arranged in 32 columns
, each parameterised by two phase
shifts $(\theta_1, \theta_2)$, plus 32 output phase shifters.  
Each one of the 32 input neurons is mixed with all 32 outputs through the
unitary transformation, making it a fully-connected layer in a single
optical pass.

Combined, the MMI tree and CMXU give FC1 a total of $336 + 1024 =
{1360}$ trainable parameters — the largest parameter block in the
entire PCNN.

\subsection{Nonlinear Optical Function Unit (NOFU)}
The 32 optical outputs of the FC1 go through a NOFU layer before passing
to the next stage. The NOFU applies a nonlinear activation function on each
of the 32 optical fields, all-optically. The design is taken directly from
Bandyopadhyay et al.~\cite{bandyopadhyay2024}, who first demonstrated this
concept on a single-chip photonic neural network.

Each NOFU unit has two trainable parameters: a power tap fraction $\beta$ and
an initial wavelength detuning $\Delta\lambda$ from the ring resonance.
Figure~\ref{fig:pool2_nofu_combined}(b) shows the circuit-level schematic of a
single unit. Light entering the NOFU first goes through a tunable coupler,
which taps off a fraction $\beta$ of the optical power to a photodiode. The
photocurrent from this photodiode is then fed back --- under a fixed bias
voltage $V_B$ --- into the carrier-injection section of a high-Q silicon
microring resonator ($Q \approx 8300$), where it generates free carriers.
These free carriers shift the resonance wavelength and introduce additional
round-trip loss, together producing a power-dependent phase and amplitude
response. The remaining $(1-\beta)$ fraction of the input field, after
passing through this now-modified ring, emerges as the nonlinearly activated
output field (Supplementary Section~II).

Figure~\ref{fig:pool2_nofu_combined}(c) shows the simulated power transmission $T$
versus input power $P_\text{in}$ for $\beta = 0.35$, $V_B = 0.8\,\text{V}$,
and three initial detunings $\Delta\lambda \in \{-0.03, 0, +0.03\}$~nm.
At low power, all three curves share a high off-resonance transmission
($T \approx 0.63$). As input power rises, carrier injection shifts the ring
resonance through the operating wavelength: the zero-detuning (green) and
$+0.03$~nm (orange) curves sweep through a deep transmission minimum around
$P_\text{in} \approx 0.7$--$2$~mW, a signature of the resonance crossing,
before recovering at high power. The $-0.03$~nm (blue) curve remains
relatively flat, as carrier injection moves the resonance further away from
the operating wavelength. Figure~\ref{fig:pool2_nofu_combined}(d) shows the
corresponding nonlinear phase shift $\Delta\Phi_\text{NL}$: all three curves
accumulate a negative phase shift reaching ${\sim}{-1.5}$~rad at the
resonance crossing, recovering toward zero once the ring is pushed far
off resonance at high power. With 32 units at 2 parameters each, the NOFU layer contributes
64 trainable parameters.

\subsection{Fully Connected Layer 2}

The 32 nonlinearly activated optical signals from the NOFU are now mapped
to 10 output classes — one per digit — through a final weighted MMI tree
followed by a $10\times10$ CMXU mesh.

The same binary 2×1 weighted MMI design from FC1 is used here, reducing
the 32 optical signals to 10 across two levels.  Level~1 reduces $32\rightarrow16$
using 16 weighted MMIs (32 phase parameters).  Level~2 reduces $16\rightarrow10$:
12 of the 16 signals are combined through 6 MMIs yielding 6 outputs, while the
remaining 4 signals pass through unchanged — together giving 10 outputs with
12 parameters.

The 10 MMI outputs enter the $10\times10$ Clements mesh. With 45 MZIs (2 phase shifts each)
and 10 output phase shifters, the mesh has ${100}$ trainable parameters,
implementing an arbitrary $10\times10$ unitary transformation: every one of the
10 input neurons is linearly mixed with all 10 outputs in a single optical pass.

After the CMXU, the 10 optical signals are converted to photocurrent by a
10-channel Ge-on-Si photodetector array, then transimpedance-amplified (TIA)
and digitised by 10-bit ADCs —
producing 10 digital logits.  The logit vector
$\mathbf{z} \in \mathbb{R}^{10}$ is passed through the softmax function to
obtain the class probability distribution:
\begin{equation}
  p_i = \frac{\exp(z_i)}{\sum_{j=0}^{9} \exp(z_j)}, \qquad i = 0, 1, \ldots, 9,
\end{equation}
and the predicted digit class is:
\begin{equation}
  \hat{y} = \arg\max_i\; p_i.
\end{equation}

This completes the full inference pipeline — from a raw $28\times28$ image
modulated onto a CW laser, through all-optical convolutional, pooling, nonlinear
and fully connected stages, to a 10-class probability vector.
The next part covers how this all-optical layers of CNN is trained.

\section{Training approach}
\label{sec:training}

Training photonic neural networks poses a unique challenge: conventional backpropagation requires access to intermediate gradients that are physically unavailable in optical circuits. Training linear as well as non-linear layers require flowing of gradients backwards. 
Unlike FICONN~\cite{bandyopadhyay2024}, which operates as a fully-connected network, our PCNN accumulates noise across convolutional, all optical pooling, WDM subsytem, MMI trees stages. Our diagnostic measurements show that the SPSA gradient SNR near the pre-trained optimum is $\approx 0.022$---making direct in-situ training from random initialisation effectively a random walk (Supplementary Section~IV). We performed this with a two-phase hybrid pipeline.

\subsection{\textit{Ex Situ} Pre-Training}

Training photonic neural networks with conventional backpropagation is physically
unrealizable because intermediate gradients are inaccessible in physical optical
circuits. To address this, we developed a PyTorch-based differentiable digital
twin that mathematically replicates every hardware operation: the unitary transfer
matrix of each MZI, the Clements rectangular mesh, the weighted MMI combining
trees, the asymmetric delay-line tap losses, the O/E/O interface,
and the carrier-injection nonlinearity of the NOFU microring resonator.

Each physical component has a direct one-to-one correspondence with the digital
twin. Because the parity between the twin and the
hardware simulator is numerically exact ($< 10^{-15}$ absolute error), the
trained parameters transfer to the hardware with a direct 1-to-1 mapping,
bypassing computationally expensive techniques like Singular Value Decomposition (SVD). 

The network was pre-trained on the MNIST dataset using the Adam optimiser~\cite{kingma2015}
with a mini-batch size of 128 images. The learning rate was modulated via a
cosine annealing schedule (Supplementary Section~S10) starting at $\eta_0 = 10^{-3}$
and decaying to $\eta_{\min} = 10^{-5}$ over 100 epochs. To ensure stability,
gradient clipping with a maximum $\ell_2$-norm of 1.0 was applied at every backward
pass. Output logits were scaled by a realistic hardware gain of $\times 1500$
and clipped to $[-15, 15]$ before the softmax operation, reflecting the dynamic
range of the on-chip TIA and ADC readout chain. Training ran on an NVIDIA A100 GPU
node in approximately
4.7~hours.

\begin{figure*}[!t]
  \centering
  \includegraphics[width=\textwidth]{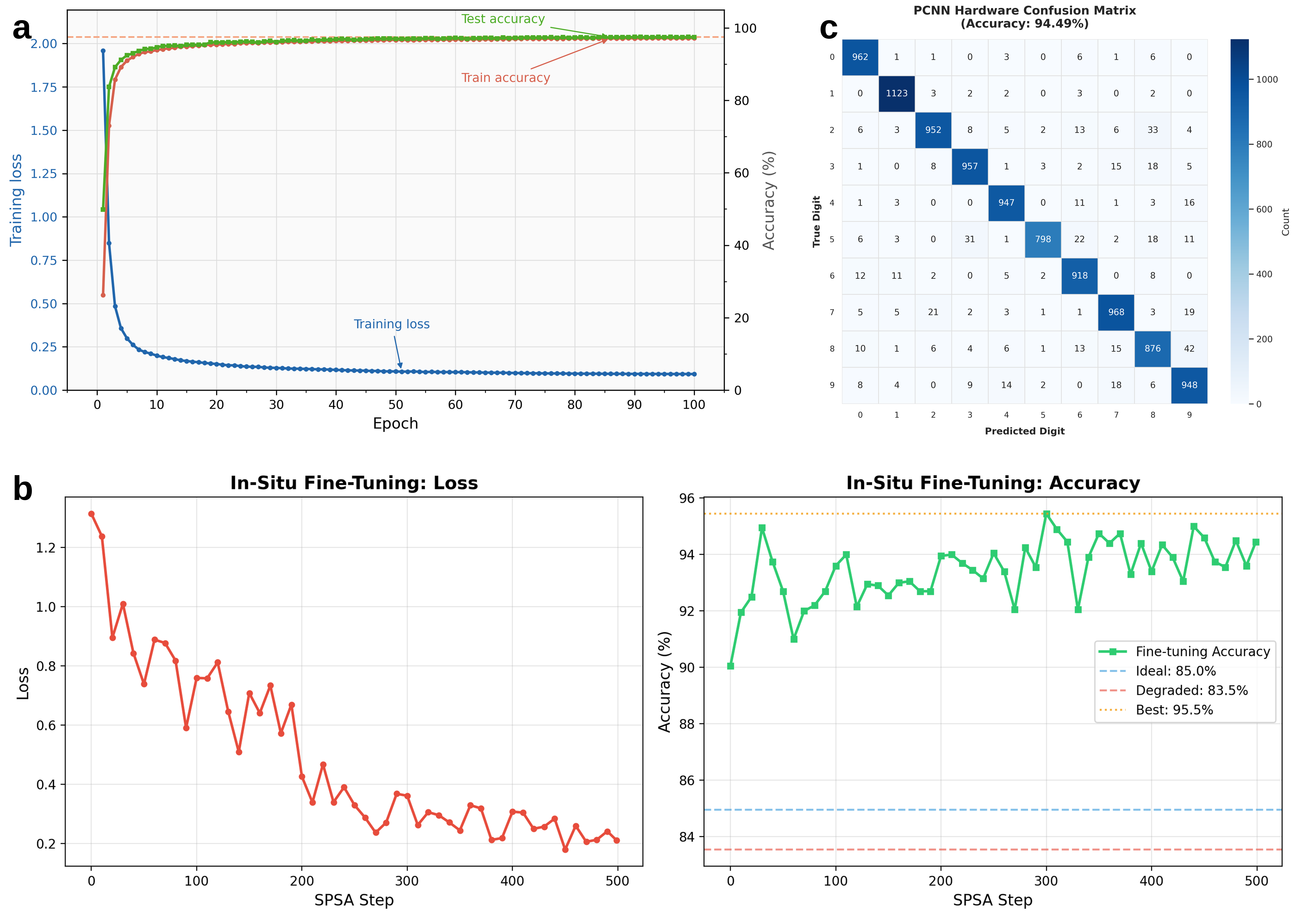}
  \caption{\textbf{PCNN training, fine-tuning, and hardware evaluation.}
  (\textbf{a})~Digital twin pre-training: training loss (blue, left axis) and
  train/test accuracy (orange/green, right axis) over 100 epochs on the
  \textsc{param Rudra} A100 cluster; test accuracy peaks at $\mathbf{97.45\%}$
  at epoch~96 with no overfitting.
  (\textbf{b})~\textit{In situ} SPSA fine-tuning: cross-entropy loss (left) and
  validation accuracy (right) vs.\ SPSA step; the best checkpoint at
  step~300 yields $\mathbf{94.49\%}$ on the full MNIST test set.
  (\textbf{c})~Hardware confusion matrix ($10{,}000$ images) after fine-tuning;
  the strong diagonal confirms uniformly high per-class accuracy across
  all ten digit classes.}
  \label{fig:training_combined}
\end{figure*}

As shown in Figure~\ref{fig:training_combined}(a), the cross-entropy training
loss decreased rapidly from 1.60 at epoch~1, converging asymptotically to $0.093$
by epoch~100. The test accuracy crossed 85\% in three epochs and peaked at
$97.45\%$ by epoch~96, with a final training accuracy of 97.17\%
confirming no overfitting.

\subsection{\textit{In Situ} Fine-Tuning}

Following \textit{ex situ} pre-training, the trained phase parameters are transferred
directly to the hardware simulation under a full non-ideality profile:
propagation loss ($\alpha = 2.0$~dB/cm), MZI insertion loss ($0.2$~dB/MZI),
fabrication phase disorder ($\sigma_\theta = 0.02$~rad,
$\sigma_\varphi = 0.05$~rad), and thermal crosstalk (coupling factor
$\xi = 0.1$). These four imperfections together degrade the transferred-weight
accuracy from $84.96\%$ on ideal hardware to $83.54\%$---a compound penalty
of $1.42\%$. To recover this loss and adapt the parameters to the physical
non-ideality environment, we apply \textit{in situ} fine-tuning directly on the
hardware simulator.

A sequential finite-difference approach to on-chip gradient estimation
requires $2N$ forward passes per step~\cite{harris2018,bandyopadhyay2024},
amounting to $3{,}592$ optical forward passes per step for our
$N=1796$ parameter model, making it impractical for on-chip training.
We instead employ Simultaneous Perturbation Stochastic Approximation
(SPSA)~\cite{spall1992}, which perturbs all parameters simultaneously
along a random direction $\boldsymbol{\Delta}$ in the parameter space,
$\boldsymbol{\Theta} \rightarrow \boldsymbol{\Theta} + \boldsymbol{\Delta}$,
and estimates the directional derivative as:
\begin{equation}
  \nabla_{\boldsymbol{\Delta}}\mathcal{L}(\boldsymbol{\Theta}) =
  \frac{\mathcal{L}(\boldsymbol{\Theta} + \boldsymbol{\Delta}) -
        \mathcal{L}(\boldsymbol{\Theta} - \boldsymbol{\Delta})}
       {2\,\|\boldsymbol{\Delta}\|}.
  \label{eq:spsa}
\end{equation}
Following standard gradient descent, the parameter vector is then updated as
$\boldsymbol{\Theta} \rightarrow \boldsymbol{\Theta} -
\eta\,\nabla_{\boldsymbol{\Delta}}\mathcal{L}(\boldsymbol{\Theta})\boldsymbol{\Delta}$,
where $\eta$ is the learning rate. Each perturbation vector
$\boldsymbol{\Delta}$ is drawn from a Bernoulli$\{-1,+1\}$ distribution.
To suppress stochastic variance, we average the estimate over $M=8$
independent perturbation draws before each update, requiring only $2M=16$
forward passes per step.
The smoothed gradient is passed to an Adam optimiser~\cite{kingma2015},
whose adaptive moment estimates provide additional noise regularisation.

Fine-tuning runs for $500$ steps with a mini-batch of $128$ randomly sampled
training images per evaluation. The Adam learning rate begins at
$5 \times 10^{-3}$ with a 50-step linear warm-up, followed by an adaptive
three-phase schedule: when no validation improvement is detected over $80$
consecutive steps, the parameter vector is rewound to the best checkpoint,
the learning rate and perturbation scale are halved, and the Adam momentum
buffers are reset. This warm-restart procedure repeats up to three times,
yielding progressively finer searches. Per-layer learning rate multipliers
further differentiate update magnitude: early convolutional layers
(Conv1: $\times 0.3$; Conv2: $\times 0.5$) are perturbed conservatively
since their features are robust to hardware drift, while output-adjacent
layers are tuned more aggressively (FC1: $\times 1.0$; NOFU: $\times 2.0$;
FC2: $\times 3.0$). 

Figure~\ref{fig:training_combined}(b) shows the convergence trajectory.
Starting from $83.54\%$, validation accuracy reaches $94.0\%$ by step~110
and climbs to $95.5\%$ following the first warm restart. The best checkpoint,
obtained at step~300, yields a final test accuracy of ${94.49\%}$ on
the full $10{,}000$-image MNIST test set---a recovery of ${+10.95\%}$
over the degraded hardware baseline---closing the gap to the digital twin
($97.45\%$) to only $2.96\%$. Figure~\ref{fig:training_combined}(c) confirms uniformly high per-class
accuracy across all ten digit classes.

\section{Discussion}

Prior demonstrations either focus on fully connected
layers only~\cite{shen2017, harris2018} or rely on electronic intermediate
processing for pooling and activation~\cite{mehrabian2018, chang2018}. While
the phase-change tensor core in~\cite{feldmann2021} achieved parallel
convolution, it required electronic max-pooling and did not support
post-programming weight reconfiguration. Lin et al.~\cite{lin2018} realised
all-optical diffractive inference, but the diffractive layers were fixed and
could not be retrained. Amiri et al.~\cite{amiri2025} demonstrated OCNN based on phase change materials with $91.90\%$ accuracy but did not have nonlinear layer. Our PCNN is the first CNN where every operation: depth-point seperable  convolution, WDM-based max-pooling, and carrier-injection
ring resonator activations---is performed entirely in the optical domain,
achieving a peak MNIST test accuracy of $94.49\%$ with only $1{,}796$ parameters.

The hybrid training pipeline demonstrates effective hardware adaptation. The adaptive warm-restart SPSA recovers accuracy from the
degraded hardware baseline and per-layer learning rate scaling is key to this recovery: conservative updates on early layers and concentrate the optimisation on output-adjacent layers.

We estimate energy efficiency from a bottom-up power budget. The $781$ MZIs
contribute $1{,}562$ thermo-optic phase shifters; at $P_\pi = 10$~mW and a mean
operating phase of $\pi/2$, the average heater dissipation is $5$~mW, yielding
a phase-shifter subtotal of $7.81$~W. The $32$ NOFU microring resonators add
$16$~mW, and the MZM modulator drivers contribute $3.0$~W, for a total static
chip power of ${10.83}$~W. Including the full electronic interface adds approximately $3.5$~W, giving a system-level estimate of $\sim$14.3~W. Inference latency is dominated by the serial
streaming of $676$ overlapping $3\times3$ patches through Conv1 at a 1~GHz
optical clock ($\tau_\mathrm{Conv1} = 676$~ns); subsequent stages are pipelined
to give $\tau_\mathrm{latency} \approx 843$~ns. The resulting single-inference
energy is ${9.1}$~$\mu$J. Across the $249{,}480$ MAC operations per
inference:
\begin{equation}
  E_\mathrm{OP} = \frac{P_\mathrm{total}\cdot\tau_\mathrm{latency}}{N_\mathrm{OPS}}
  = \frac{10.83~\text{W} \times 843~\text{ns}}{249{,}480} \approx 36.5~\text{pJ/OP}.
\end{equation}
This places the PCNN ${220}$--${330}\times$
more energy-efficient than state-of-the-art electronic GPUs for single-image
inference. While the $0.32$~TOPS throughput is limited by the serial Conv1
streaming bottleneck, upgrading to undercut thermal heaters~\cite{wright2022}
($P_\pi = 3$~mW) or MEMS-based actuators~\cite{baghdadi2021,gyger2021}
($P_\pi = 10~\mu$W) would reduce the phase-shifter subtotal to $2.34$~W, lowering $E_\mathrm{OP}$ to $\sim$18~pJ/OP and $\sim$10~pJ/OP
respectively, approaching the fundamental limits of optical computation.


Scaling to more complex datasets (e.g., CIFAR-10 or ImageNet) requires
deepening the convolutional hierarchy and expanding the MZI mesh dimensions,
increasing chip area and phase control complexity. The serial streaming
bottleneck can be mitigated by WDM-parallel patch encoding or spatial-mode
multiplexing at the modulator array. Because our analysis relies entirely on
numerical hardware models, physical fabrication and measurement are required
to characterise wafer-level spatial variations.

\section{Conclusion}

We have demonstrated a fully integrated photonic convolutional neural network
on a single silicon photonic chip of estimated footprint $18 \times 18$~mm$^2$,
capable of performing MNIST digit classification in the optical domain
at a peak test accuracy of $94.49\%$ with $1{,}796$ tunable parameters.
Unlike prior hybrid architectures that intersperse optical computation with
electronic readout at every layer, the PCNN performs convolution, max-pooling,
and nonlinear activation coherently in the optical domain throughout, with a
single O/E/O interface only before the first fully connected layer,
where all $200$ optically computed feature values must be collected
simultaneously from the $25 \times 8$ WDM output of the second pooling stage. This design significantly reduces
the latency and power overhead that electronic conversion stages would otherwise
introduce between every layer. The training methodology introduced here combines \textit{ex situ} backpropagation with \textit{in situ} SPSA fine-tuning
directly on the hardware simulator. The \textit{in situ} SPSA stage then
adapted these parameters to the specific non-ideality environment---propagation
loss, MZI insertion loss, fabrication phase disorder, and thermal crosstalk---
using only $2M = 16$ hardware forward passes per step. This decouples the training 
cost from the model parameter count, requiring $\mathcal{O}(1)$ hardware passes 
per step with respect to the number of parameters $N$—a critical advantage over 
sequential finite-difference methods which perturbs the parameters one by one, requiring $\mathcal{O}(N)$ passes per step 
(amounting to $2N = 3{,}592$ passes in our case). For a complete training run of 
$S$ optimization steps, the total hardware complexity scales as $\mathcal{O}(M \cdot S)$ 
rather than $\mathcal{O}(N \cdot S)$, ensuring excellent scalability to larger, 
more complex photonic architectures.

Because SPSA requires
only two optically accelerated forward passes per gradient estimate, a
fabricated PCNN could in principle run on-chip fine-tuning in real time,
correcting for slow thermal drift or phase offsets without
interrupting operation. This capability is directly relevant to edge-inference
scenarios such as LiDAR-based perception in autonomous vehicles and UAV
platforms, where a compact photonic processor performing image sensing,
feature extraction, and classification within a single integrated chip would
offer decisive advantages in latency, power, and form factor over conventional
electronic solutions. Physical fabrication of the $18 \times 18$~mm$^2$
silicon photonic chip is the natural next step, and the numerical models
developed here---the hardware simulator, the digital twin, and
the fine-tuning pipeline---provide a complete design and framework.

\section*{Data availability}
The MNIST handwritten digit dataset used for training and evaluation is publicly available at \url{http://yann.lecun.com/exdb/mnist/}. All numerical simulation results reported in this work are derived entirely from software models; no experimental raw data was generated.

\section*{Code availability}
The source code, PyTorch digital twin model, and hardware simulation frameworks (including the SPSA in-situ fine-tuning pipeline) developed in this work are available from the corresponding author (S.R.) upon reasonable request.

\section*{Acknowledgements}
We thank the Inter-University Accelerator Centre (IUAC), New Delhi, for access to the \textsc{param Rudra} high-performance computing facility (NVIDIA A100 GPU nodes), on which all pre-training experiments reported in this work were performed. We thank Adam Darcie for his valuable insights on hardware non-idealities, and Shikha Gupta and Xavier-Lewis Palmer for their informal guidance and constructive discussions.

\section*{Author contributions}
S.R.\ conceptualised the PCNN, designed the full architecture, implemented the simulation codebase, built the training and debugging pipeline, and drafted the manuscript. S.T.\ and A.S.\ critically reviewed the project and provided guidance throughout the process.

\section*{Competing interests}
The authors declare no competing interests.

\section*{Additional information}
\textbf{Supplementary information} The online version contains supplementary material.

\noindent\textbf{Correspondence} and requests for materials should be addressed to Saurabh Ranjan.

\bibliographystyle{unsrt}

\end{document}


\maketitle
\vspace{-0.5em}
{\centering\large\bfseries Contents\par}
\vspace{0.5em}
{\centering
\begin{tabular}{@{}p{0.70\textwidth}r@{}}
\textbf{Section} & \textbf{Page} \\
\midrule
I. Device Characterization & \pageref{sec:device_characterization} \\
\quad Clements Mesh & \pageref{sec:clements_mesh} \\
\quad Convolution Layer 2 & \pageref{sec:conv_layer_2} \\
\quad Fully Connected Layer~1 & \pageref{sec:fc_layer_1} \\
II. NOFU & \pageref{sec:s3} \\
III. WDM-based Pooling & \pageref{sec:wdm_pooling} \\
IV. Training & \pageref{sec:training} \\
\quad Signal-to-Noise Ratio Analysis & \pageref{sec:snr_analysis} \\
\quad Pre-training & \pageref{sec:pretraining} \\
V. Latency and energy efficiency & \pageref{sec:latency_energy} \\
VI. Scaling & \pageref{sec:scaling} \\
\bottomrule
\end{tabular}\par}
\clearpage
\section{Device Characterization}
\label{sec:device_characterization}

\subsection{Clements Mesh}
\label{sec:clements_mesh}

We implement arbitrary $N \times N$ unitary transformations using the rectangular Clements mesh configuration~\cite{clements2016}, arranging $N(N-1)/2$ Mach-Zehnder interferometers (MZIs) in a symmetric crossing pattern. We parameterize each MZI using an internal phase shift $\theta_1$ to control the splitting ratio and an external phase shift $\theta_2$ to apply a differential phase. 

\begin{figure}[H]
  \centering
  \includegraphics[width=0.55\textwidth]{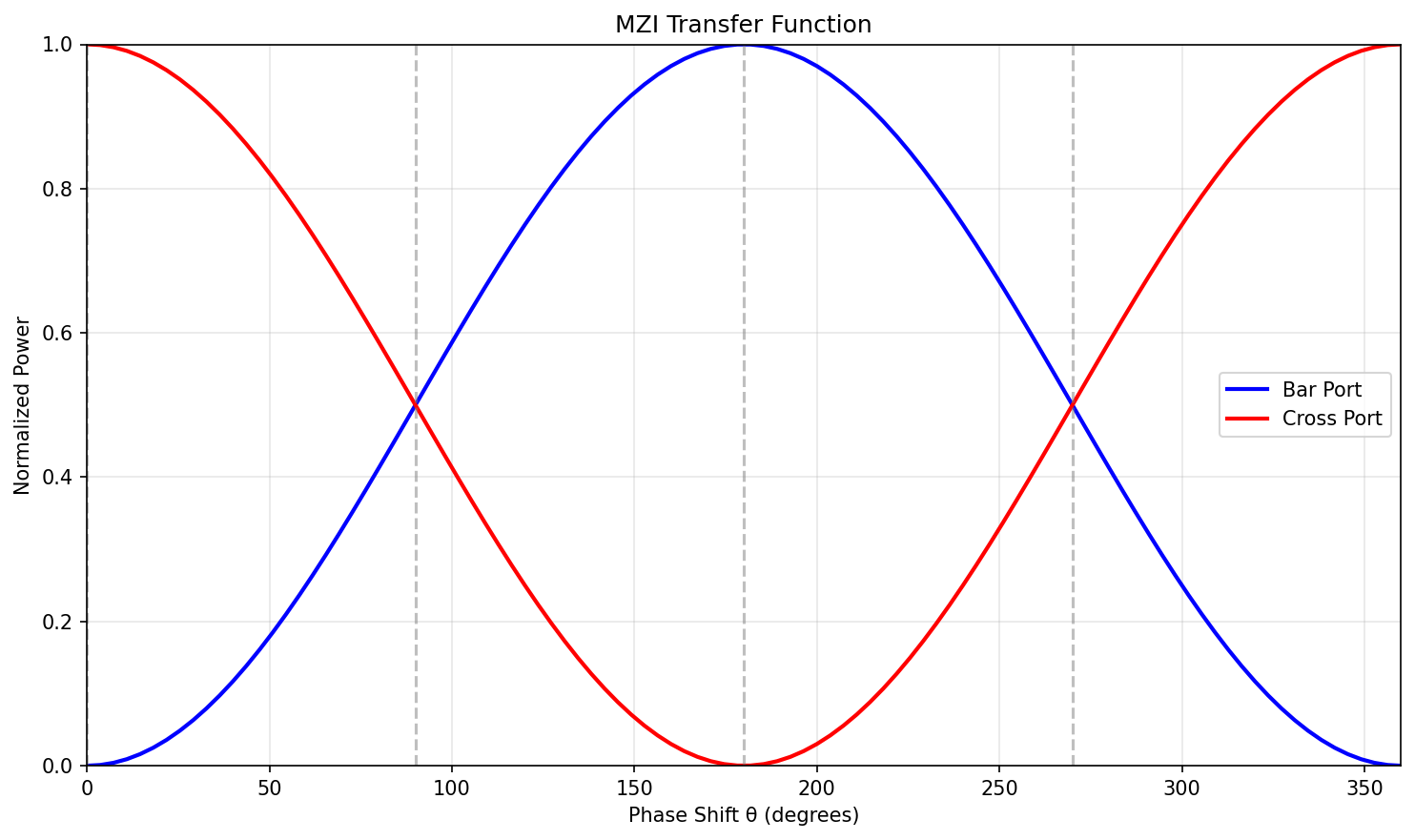}
  \caption{\textbf{MZI transfer function.} Simulated optical transmission through the bar (blue) and cross (red) ports as a function of the internal phase shift $\theta_1$.}
  \label{fig:mzi_transfer}
\end{figure}

To characterize a single MZI, we launch light into a single input port and measure the output transmission $T(\theta_1) = P_{\text{out}}/P_{\text{in}}$. For an ideal device, the transmission at the bar and cross ports scales as:
\begin{equation}
T_{\text{bar}}(\theta_1) = \sin^2\left(\frac{\theta_1}{2}\right)
\label{eq:tbar}
\end{equation}
and at the cross port:
\begin{equation}
T_{\text{cross}}(\theta_1) = \cos^2\left(\frac{\theta_1}{2}\right).
\label{eq:tcross}
\end{equation}
In our PCNN, the bar and cross ports control the spatial routing and interference of optical fields between adjacent waveguide channels to implement the programmed weight matrices, as shown in Figure~\ref{fig:mzi_transfer}. Each MZI implements the $2 \times 2$ unitary matrix:
\begin{equation}
U(\theta_1, \theta_2) = i\, e^{i\theta_1/2}
\begin{pmatrix}
e^{i\theta_2}\sin(\theta_1/2) & e^{i\theta_2}\cos(\theta_1/2) \\
\cos(\theta_1/2) & -\sin(\theta_1/2)
\end{pmatrix}.
\end{equation}
For every Clements mesh instance in our PCNN, we verified that the realized transformation satisfies the unitary condition $U^\dagger U = I$.

\subsection{Convolution Layer 2}
\label{sec:conv_layer_2}

\begin{figure}[H]
  \centering
  \includegraphics[width=\textwidth]{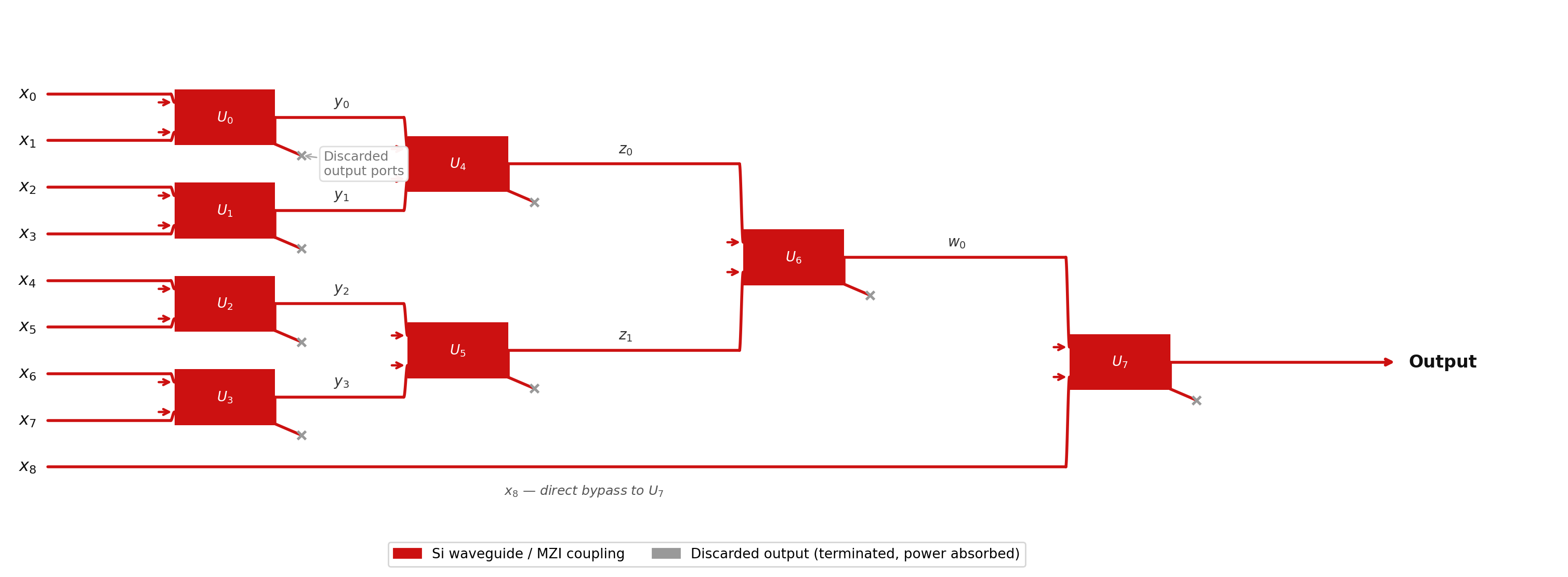}
  \caption{\textbf{$9\rightarrow1$ MZI binary combiner tree for Conv2 Step~2A (depthwise convolution).}
  Nine $3\times3$ patch inputs ($x_0$--$x_8$) are coherently combined into a 
  single output through four tree levels using 8 Mach--Zehnder interferometers. Terminated ($\times$) stubs denote discarded output ports whose power is
  absorbed; all useful optical power is routed to the single output port.
  Four identical trees operate in parallel (one per input channel) for a total
  depthwise count of 32~MZIs and 64~phase parameters.}
  \label{fig:conv2a_mzi_tree}
\end{figure}

Each of the 4 input channels to Conv2 Step~2A arrives as a serial stream of
169 optical field samples at 1~GHz, representing the $13\times13$ feature map. A $1\times9$ MMI splitter fans out this stream into 9
parallel waveguides, each routed through a spiral silicon nitride delay line of
a distinct length, into the MZI combiner tree (schematic shown in Fig.~\ref{fig:conv2a_mzi_tree}).

The general formula for path $p$ (0-indexed, $p \in \{0,\ldots,8\}$) is
\begin{equation}
  \Delta t_p = (8 - p) + \left\lfloor \frac{8-p}{3} \right\rfloor \times (W - 3),
  \label{eq:delay_formula}
\end{equation}
where the first term accounts for the column offset within the $3\times3$
window and the second term for the additional inter-row delay of $(W-3)$
clock periods.

Conv2 Step~2B applies a pointwise convolution using a dedicated $8\times8$ Clements mesh that implements an arbitrary $8\times8$ unitary transformation $U^{(8)}$.

\subsection{Fully Connected Layer~1}
\label{sec:fc_layer_1}

The FC~1 stage requires all 200~values simultaneously as optical signals at a single wavelength~$\lambda_0$, but bridging this serial-to-parallel conversion requires a three-component opto-electro-opto (O/E/O) interface.

In the O/E conversion stage, eight Ge-on-Si photodetectors ($20\,\mu\text{m}\times20\,\mu\text{m}$, responsivity 1~A/W, bandwidth $>10$~GHz) sample the eight wavelength channels in parallel. Each photocurrent is converted to a 0--1~V analog voltage by a transimpedance amplifier (TIA, gain 1~k$\Omega$), and then digitised by a 10-bit ADC running at 1~GS/s at one sample per nanosecond. The digital values stream into an electronic shift-register buffer consisting of 8~shift registers, one per channel, each 25~stages deep. Every stage is a 10-bit D-flip-flop, giving $200\times10 = 2{,}000$~flip-flops in total, indexed as:
\begin{equation}
  \text{Reg}\!\left[25(c-1)+p\right] \;\leftarrow\; \text{val}^{(c)}[p],
  \qquad c\in\{1,\ldots,8\},\;\; p\in\{0,\ldots,24\},
\end{equation}
such that channel~$c$ occupies registers $25(c-1)$ through $25c-1$. Table~\ref{tab:buffer_timing} summarises the buffer filling timeline: Pool~2's first output arrives at $t \approx 330$~ns, with one new 8-sample row filling the buffer every nanosecond until all 200~registers are stable at $t = 355$~ns, triggering the HOLD phase.

\begin{table}[H]
\centering
\begin{tabular}{@{}llll@{}}
\toprule
\textbf{Time} & \textbf{Event} & \textbf{Registers filled}
  & \textbf{Buffer state} \\ \midrule
$t = 330$~ns & Pool2 output starts & R[0], R[25],\ldots,R[175] & 8 / 200 \\
$t = 331$~ns & 2nd sample & R[1], R[26],\ldots,R[176] & 16 / 200 \\
$t = 332$~ns & 3rd sample & R[2], R[27],\ldots,R[177] & 24 / 200 \\
$\vdots$ & $\vdots$ & $\vdots$ & $\vdots$ \\
$t = 354$~ns & 25th sample & R[24], R[49],\ldots,R[199] & 200 / 200 \\
$t = 355$~ns & HOLD phase begins & all 200 stable & ready \\
$t = 355\text{--}356$~ns & Parallel readout → DAC → MZM & all 200 simultaneously & E→O \\
\bottomrule
\end{tabular}
\caption{FC1 buffer filling timeline. Pool~2 first output arrives at
         $t\approx330$~ns from the start of frame processing.}
\label{tab:buffer_timing}
\end{table}

Once the HOLD phase begins, all 200 digital values are simultaneously converted to analog by 200 individual DACs, each driving a dedicated Mach-Zehnder modulator (MZM, extinction ratio $>20$~dB) that re-modulates a fraction of the shared CW laser at $\lambda_0 = 1550$~nm. The modulator array is arranged as $20\,\text{rows}\times10\,\text{columns}$, and all 200~MZMs switch simultaneously, completing the entire serial-to-parallel conversion in a single clock cycle ($\sim$1~ns).

The resulting 200~parallel optical signals are then reduced to 32 via a three-level binary MMI tree. Each node is a trainable $2\times1$ weighted MMI whose two input arms each carry a thermo-optic phase shifter ($\phi_1$, $\phi_2$), giving an output field:
\begin{equation}
  E_\text{out} = \frac{E_1 e^{i\phi_1} + E_2 e^{i\phi_2}}{\sqrt{2}},
\end{equation}
and after photodetection the intensity is:
\begin{equation}
  |E_\text{out}|^2 = \frac{|E_1|^2 + |E_2|^2 + 2|E_1||E_2|\cos(\Delta\phi)}{2},
\end{equation}
where $\Delta\phi = \phi_1 - \phi_2$ is the trainable relative phase that controls the effective weighting between the two inputs.


\section{NOFU}

\label{sec:s3}

We adopt the NOFU design from Bandyopadhyay et al.~\cite{bandyo2022}. Each of the 32 NOFU units uses a photodiode feedback loop to implement a power-dependent optical nonlinearity. A tap photodiode measures the circulating power and drives forward-biased carrier injection into the ring, which shifts the resonance and increases round-trip loss. The empirical fits to the measured device data are:
\begin{align}
  \Delta\phi_{\text{NL}} &= -1.18 \times I_{\text{mA}}^{0.746},
  \label{eq:nofu_phase} \\
  a_{\text{NL}} &= \exp\!\left(-0.08 \times I_{\text{mA}}^{0.611}\right),
  \label{eq:nofu_loss}
\end{align}
where $\Delta\phi_{\text{NL}}$ (rad) is the carrier-induced resonance blueshift and $a_{\text{NL}}$ is the fractional amplitude reduction per round trip due to free-carrier absorption.

At low incident power the ring is overcoupled, and circulating power is high. As input power rises, carrier injection increases the round-trip loss and drives the ring toward the undercoupled regime, reducing cavity buildup. This power-dependent coupling transition produces the sigmoidal-like output, tunable via the feedback gain $\beta$ and the operating detuning $\Delta\lambda$. At low power, off-resonance transmission is high ($T \approx 0.63$); as power rises the carrier-induced resonance shift sweeps through the operating wavelength, driving $T$ through a deep minimum, then recovering once the ring is pushed far off resonance.


\section{WDM-based Pooling}

\label{sec:wdm_pooling}

Spiral delay networks which dominates the pooling area, aligns the spatial values with each delay lines represent each feature channels. Scaling this optical network linearly as N(number of channels) grows is infeasible for deeper networks. To address this problem we assign feature channels to a distinct wavelength and combined all wavelmngth representing channels onto a single waveguide through Arrayed Waveguide Grating(AWG)~\cite{zou2022} combiner.  

The combined multi-wavelength signal then traverses a single shared delay network of 4 spiral delay lines, after which AWG demultiplexers separate the channels for per-channel max-pool comparison. Table~\ref{tab:area_component} breaks down silicon area at $N{=}8$ channels, while Table~\ref{tab:area_scaling} shows how total area scales with channel count.

\begin{table}[H]
\centering
\begin{tabular}{@{}lcc@{}}
\toprule
\textbf{Component} & \textbf{Conventional} & \textbf{WDM-based} \\ \midrule
Delay networks & 72~mm$^2$ ($N\times9$) & 9~mm$^2$ ($1\times9$) \\
AWG combiner / demux & --- & 5~mm$^2$ \\
Max-pool units & 0.56~mm$^2$ & 0.56~mm$^2$ \\
Routing / misc. & 3~mm$^2$ & 3~mm$^2$ \\ \midrule
\textbf{Total} & \textbf{$\sim$81~mm$^2$} & \textbf{$\sim$19~mm$^2$} \\
\bottomrule
\end{tabular}
\caption{Pool2 component-level area at $N=8$: conventional vs.\ WDM.}
\label{tab:area_component}
\end{table}

\begin{table}[H]
\centering
\begin{tabular}{@{}lcccc@{}}
\toprule
\textbf{Channels ($N$)} & \textbf{Conventional} & \textbf{WDM-based} & \textbf{Savings} & \textbf{Reduction} \\ \midrule
4  & $\sim$40~mm$^2$ & $\sim$32~mm$^2$ & 8~mm$^2$   & 20\% \\
8  & $\sim$81~mm$^2$ & $\sim$19~mm$^2$ & 62~mm$^2$  & 76\% \\
16 & $\sim$162~mm$^2$& $\sim$26~mm$^2$ & 136~mm$^2$ & 84\% \\
32 & $\sim$324~mm$^2$& $\sim$40~mm$^2$ & 284~mm$^2$ & 88\% \\ \bottomrule
\end{tabular}
\caption{Total area scaling with channel count $N$. Conventional scales as $\mathcal{O}(N)$; WDM scales as $\mathcal{O}(1)$ (only AWG/pool units grow).}
\label{tab:area_scaling}
\end{table}

We selectively apply WDM only at Pool~2 because each AWG pair carries a fixed overhead of $\approx5$~mm$^2$ plus $\sim1$--2~dB insertion loss. For earlier stages with fewer channels---Pool~1 ($N{=}4$) and the Conv~2 depthwise step ($N{=}4$ MZI trees)---this fixed overhead offsets the delay-line savings entirely, as the $N{=}4$ row of Table~\ref{tab:area_scaling} confirms (only 20\% savings, versus the cost of the AWG pair). Pool~2 is the first stage in the PCNN where the WDM overhead is decisively outweighed by the area and power benefits, making it the natural crossover point above which WDM becomes the right architectural choice.

\section{Training}

\label{sec:training}

\subsection{Signal-to-Noise Ratio Analysis}
\label{sec:snr_analysis}
To verify that direct in-situ SPSA training from random initialisation is not viable for the PCNN, we measured the SPSA gradient SNR. We ran $n = 5$ independent gradient estimates $\{\hat{g}_k\}_{k=1}^{5}$ at the pre-trained parameter point $\boldsymbol{\Theta}^*$, each on a mini-batch of 5 training images, and computed:
\begin{equation}
  \mathrm{SNR} = \frac{|\mathbb{E}[\hat{g}]|}{\mathrm{std}(\hat{g})}.
\end{equation}

The mean SPSA gradient SNR for the PCNN across all 1,796 parameters is $\approx 0.022$. At SNR~$\approx 0.022$, the SPSA update is dominated by stochastic noise and is effectively indistinguishable from a random walk. Even averaging over $M = 8$ perturbation draws improves the effective SNR to only $0.022 \times \sqrt{8} \approx 0.062$.
It is this combination---high-dimensional parameter space, multi-stage noise accumulation that makes convergence from random initialisation impractical for the PCNN. 

\subsection{Pre-training}
\label{sec:pretraining}

To optimize pre-training on NVIDIA A100 GPU, we evaluated mini-batch sizes of 32, 128, and 512. Although a batch size of 32 is a typical choice, the waveguide-by-waveguide propagation inside the Clements mesh in complex128 precision is CPU-bound. With a batch size of 32, the A100 GPU remained mostly idle, leading to an impractical throughput of under 5 samples per second (projecting to over 7 hours per epoch). Batch size 128 resolved this bottleneck, providing full GPU utilization and completing one epoch in approximately 5 minutes (50 epochs in 2.5 hours), while maintaining the gradient update frequency required for stable convergence.

\begin{figure}[H]
  \centering
  \includegraphics[width=\textwidth]{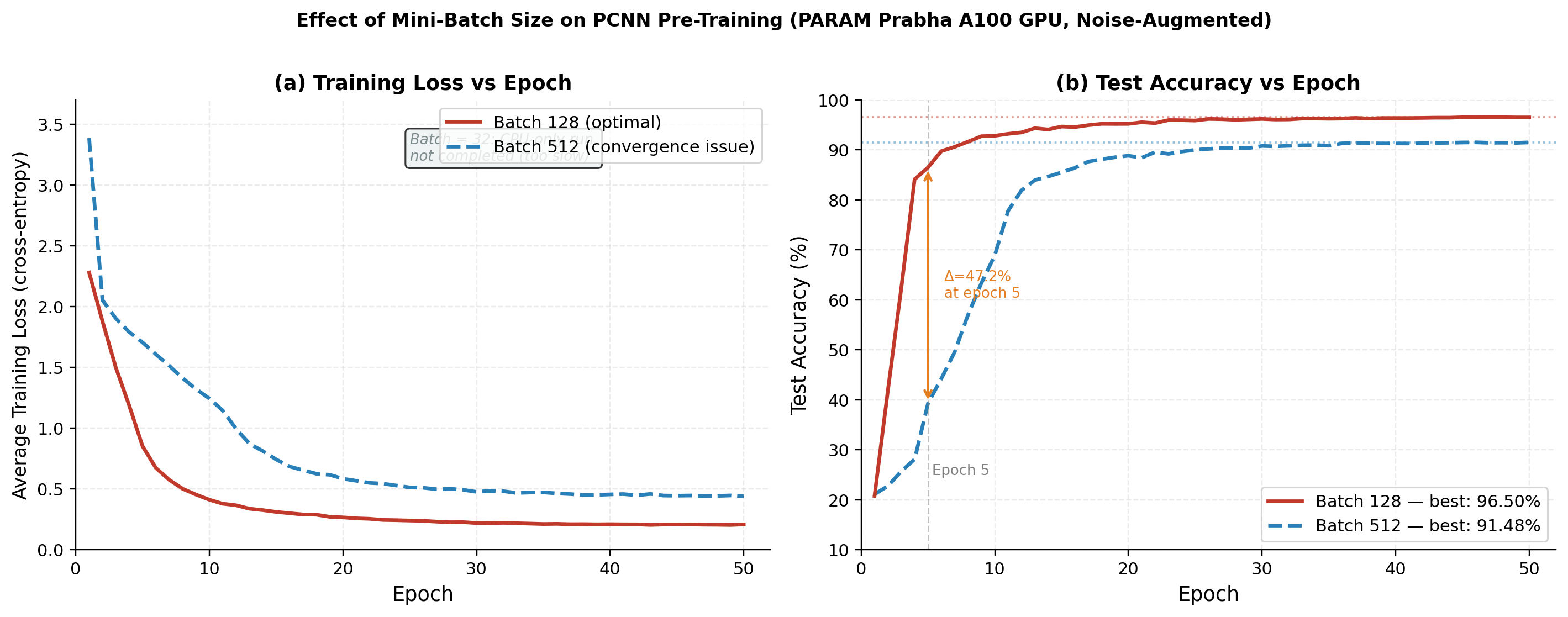}
  \caption{\textbf{Effect of mini-batch size on PCNN pre-training}
  (\textit{a})~Training loss vs.\ epoch: batch~128 converges to 0.21
  while batch~512 stagnates above 0.44.
  (\textit{b})~Test accuracy vs.\ epoch: at epoch~5, batch~128 already
  achieves 86.5\% while batch~512 reaches only 39.3\%.  Final best accuracy is 96.50\% (batch~128)
  vs.\ 91.48\% (batch~512).  Batch~32 was terminated early due to
  GPU under-utilisation on the A100 hardware.}
  \label{fig:batchsize}
\end{figure}

While batch size 512 maximized parallel throughput, it significantly degraded model convergence compared to batch 128, as shown in Figure~\ref{fig:batchsize}. At epoch 5, batch 512 achieved only 39.3\% test accuracy compared to 86.5\% for batch 128, and eventually asymptoted at 91.48\% while batch 128 reached 96.50\%. This degradation is caused by the large-batch generalization penalty~\cite{keskar2017}, where the lower variance of larger batches causes the optimizer to settle into sharp, poorly generalizing local minima. We therefore adopted a mini-batch size of 128 as the standard configuration for all reported pre-training runs.

\section{Latency and energy efficiency}

\label{sec:latency_energy}

We perform a bottom-up power estimate from the component inventory of our PCNN hardware (Table~\ref{tab:energy_params}). The thermo-optic phase shifters dominate the static power budget: 1,562 heaters (two per MZI, $N_\mathrm{MZI} = 781$) each dissipate 5~mW at mean operating phase $\pi/2$ ($P_\pi = 10$~mW), giving a subtotal of 7.81~W~\cite{harris2018}. The 32 NOFU microring resonators add 16~mW, and the 201 MZM drivers (1 input encoder $+$ 200 re-modulation units) contribute 3.0~W~\cite{tait2017}, for a total \textit{on-chip} photonic power of $P_\mathrm{chip} = 10.83$~W. The O/E/O interface electronics---8 TIAs, 8 ADCs, 200 weight DACs, the 2,000-flip-flop digital shift-register buffer, and the CW laser source---contribute a further $\approx 3.3$~W~\cite{bandyopadhyay2024}, giving a full system-level estimate of $\sim 14.1$W; the figure reported in the main text additionally accounts for PCB routing and voltage regulator overhead.

Inference latency is dominated by the serial streaming of 676 overlapping $3{\times}3$ patches through Conv1 at 1~GHz ($\tau_\mathrm{Conv1} = 676$~ns); all subsequent layers are pipelined, adding $\approx 167$~ns, giving $\tau_\mathrm{latency} = 843$~ns. Energy per inference is $E = P_\mathrm{chip} \times \tau_\mathrm{latency} = 9.1~\mu$J. With $N_\mathrm{OPS} = 249{,}480$ MAC operations , the energy per operation is $E_\mathrm{OP} = 36.5$~pJ/OP---220--330$\times$ lower than state-of-the-art NVIDIA GPUs.

\begin{table}[H]
\centering
\begin{tabular}{@{}lrl@{}}
\toprule
\textbf{Parameter} & \textbf{Value} & \textbf{Reference} \\ \midrule
\multicolumn{3}{@{}l}{\textit{On-chip photonic components}} \\
\quad Phase shifter $P_\pi$ (TiN thermo-optic) & 10~mW & \cite{harris2018} \\
\quad Phase shifter $P_\mathrm{avg}$ (mean bias $\pi/2$) & 5~mW & derived \\
\quad Thermo-optic heaters ($2 \times N_\mathrm{MZI}$) & 1,562 & verified \\
\quad Phase-shifter subtotal & 7.81~W & derived \\
\quad NOFU microring (carrier injection) & 0.5~mW each & \cite{bandyopadhyay2024} \\
\quad NOFU subtotal & 16~mW & derived \\
\quad MZM driver (input + re-modulation) & 15~mW each & \cite{tait2017} \\
\quad MZM subtotal & 3.0~W & derived \\
\quad \textbf{Total on-chip power} $P_\mathrm{chip}$ & \textbf{10.83~W} & derived \\ \midrule
\multicolumn{3}{@{}l}{\textit{O/E/O interface electronics (1~GHz)}} \\
\quad Transimpedance amplifier (TIA, receiver) & 50~mW each & \cite{bandyopadhyay2024} \\
\quad TIA subtotal & 400~mW & derived \\
\quad ADC (10-bit, 1~GS/s, receiver) & 100~mW each & \cite{bandyopadhyay2024} \\
\quad ADC subtotal & 800~mW & derived \\
\quad DAC (weight encoding, static hold) & 5~mW each & derived \\
\quad DAC subtotal & 1.0~W & derived \\
\quad Digital shift-register buffer (2,000 FFs at 1~GHz) & --- & --- \\
\quad Buffer subtotal & $\approx$1.0~W & derived \\
\quad CW laser source & --- & --- \\
\quad Laser subtotal & $\approx$50~mW & derived \\
\quad \textbf{Total interface power} & \textbf{$\approx$~3.3~W} & derived \\ \midrule
\bottomrule
\end{tabular}
\caption{Parameters used for energy efficiency calculation. On-chip photonic values are verified against the hardware implementation. }
\label{tab:energy_params}
\end{table}

\begin{table}[H]
\centering
\begin{tabular}{@{}lrrr@{}}
\toprule
\textbf{Phase shifter} & \textbf{Power} & \textbf{Energy/op} & \textbf{Ratio} \\ \midrule
\textbf{Thermal (this work)} & \textbf{10.83~W}  & \textbf{36.5~pJ/OP} & \textbf{1$\times$} \\
Undercut thermal~\cite{wright2022} & 5.36~W & 18.1~pJ/OP & 0.49$\times$ \\
MEMS~\cite{baghdadi2021,gyger2021} & 3.02~W & 10.2~pJ/OP & 0.28$\times$ \\

\bottomrule
\end{tabular}
\caption{Energy comparison under alternative phase-shifter technologies.}
\label{tab:energy_si}
\end{table}

\section{Scaling}

\label{sec:scaling}

Extending the PCNN to CIFAR-10 or ImageNet requires multi-channel RGB input, deeper convolutional stages, and larger MZI meshes. Scaling a passive optical network, however, introduces physical limits: cumulative insertion and propagation loss grows with network depth. Maintaining an acceptable SNR at larger scale will require either higher launch powers or on-chip optical gain to re-amplify signals between stages.

The current PCNN is well matched to edge inference where power and weight are hard constraints. In unmanned aerial vehicles, electronic GPUs are too power-hungry to deploy. The PCNN processes visual data for obstacle avoidance and navigation at a fraction of the energy cost. Coherent photonic meshes also do not rely on switching digital CMOS logic, which makes them more tolerant to high-radiation environments than electronic processors. In autonomous driving, near-zero-latency convolution at the speed of light is directly useful for real-time object detection. In space exploration rovers, strict battery and payload limits favor the deployment of photonic CNN systems. These systems can process multi-dimensional data with low power dissipation, enabling real-time object detection, collision avoidance, and autonomous maneuvering.